\documentclass[aps,nofootinbib,floatfix,showpacs,preprintnumbers,amsmath,amssymb]{revtex4}
\usepackage{natbib}
\usepackage{dcolumn}
\usepackage{graphicx}
\usepackage{enumerate}
\usepackage{inputenc}
\usepackage{fontenc}
\usepackage{float}
\usepackage{graphicx, rotating, lscape, threeparttable}% \usepackage{amsmath}
\usepackage{xcolor}
\usepackage{multirow}
\usepackage{bm}
\usepackage[colorlinks]{hyperref}
\usepackage{changepage}
\newcommand{\be}{\begin{equation}}
\newcommand{\ee}{\end{equation}}
\newcommand{\bea}{\begin{eqnarray}}
\newcommand{\eea}{\end{eqnarray}}

\def\[{\begin{equation}}
\def\]{\end{equation}}

\begin{document}
\title{Cosmological discordances III: more on measure properties, Large-Scale-Structure constraints, the Hubble constant and Planck}
\date{\today}
\author{Cristhian Garcia-Quintero$^1$}
\email{gqcristhian@utdallas.edu}
\author{Mustapha Ishak$^1$}
\email{mishak@utdallas.edu}
\author{Logan Fox$^1$}
\email{lkf140030@utdallas.edu}
\author{Weikang Lin$^2$}
\email{wlin23@ncsu.edu}

\affiliation{$^1$Department of Physics, The University of Texas at Dallas, Richardson, Texas 75080, USA}
\affiliation{$^2$Department of Physics, North Carolina State University, Raleigh, NC 27695, USA}
\begin{abstract}
Consistency between cosmological data sets is essential for ongoing and future cosmological analyses. We first investigate the questions of stability and applicability of some moment-based inconsistency measures to multiple data sets. We show that the recently introduced index of inconsistency (IOI) is numerically stable while it can be applied to multiple data sets. We use an illustrative construction of constraints as well as an example with real data sets (i.e. WMAP versus Planck) to show some limitations of the application of the Karhunen-Loeve decomposition to discordance measures. Second, we perform various consistency analyzes using IOI between multiple current data sets while \textit{working with the entire common parameter spaces}. We find current Large-Scale-Structure (LSS)  data sets (Planck CMB lensing, DES lensing-clustering and SDSS RSD) all to be consistent with one another. This is found to be not the case for Planck temperature (TT) versus polarization (TE,EE) data, where moderate inconsistencies are present. 
Noteworthy, we find a strong inconsistency between joint LSS probes and Planck with IOI=5.27, and a moderate tension between DES and Planck with IOI=3.14. Next, using the IOI metric, we compare the Hubble constant from five independent probes. We confirm previous strong tensions between local measurement (SH0ES) and Planck as well as between H0LiCOW and Planck, but also find new strong tensions between SH0ES measurement and the joint LSS probes with IOI=6.73 (i.e. 3.7-$\sigma$ in 1D) as well as between joint LSS and combined probes SH0ES+H0LiCOW with IOI=8.59 (i.e. 4.1-$\sigma$ in 1D).
Whether due to systematic effects in the data sets or problems with the underlying model, sources of these old and new tensions need to be identified and dealt with.
\end{abstract}
\maketitle
%\newpage
\setcounter{tocdepth}{2}
%\tableofcontents
%\newpage

%%%%%%%%%%%%%%%%%%%%%%%%%%%%%%%%%%%%%%%%%%%%%%%%%

%%%%%%%%%%%%%%%%%%%%%%%%%%%%%%%%%%%%%%%%%%%%%%%%%
\section{introduction}
%%%%%%%%%%%%%%%%%%%%%%%%%%%%%%%%%%%%%%%%%%%%%%%%%%%%
%
The concordance cosmological $\Lambda$CDM-model has been an incredible achievement of modern cosmology for at least two decades, as it has been overall a very good fit to all types of cosmological and astrophysical data sets. However, persistent inconsistencies between some cosmological parameters have started challenging the serenity of this concordance. Such inconsistencies can be due to problems with the underlying model, or due to some systematic errors in the data. 

At the forefront of such inconsistencies comes the persistent tension between the local measurements of the Hubble constant, $H_0$, from the
HST SH0ES measurements based on the Cepheid distance scale and the value inferred from the CMB Planck data. This tension has now risen to the 4.4-$\sigma$ level \cite{2019-Riess-H0}, and has attracted much work and debate in the literature \cite{Poulin-etal-2019-EDE,2014Efstathiou-Hubble,2019-Riess-H0,Freedman-etal-2019,Yuan-etal-2019-TRGB-local,2019-Arendse-etal,2019-Knox-Millea,2019-Davis-etal,2016-Valentino-Melchiorri-Silk-reconciling-Planck-local,2019-Kenworthy-Scolnic-Riess,2018-Pustilnik-etal-local-void}. Several different methods of calibration have been used in the SH0ES local measurements and they all lead to significantly higher values of $H_0$ than the ones obtained from Planck. This argues against attributing the tension to systematics in the local measurements. Most recently, Ref. \cite{Freedman-etal-2019} used a calibration of the Tip of the Red Giant Branch (TRGB) applied to Type Ia supernovae and derived a value of  $H_0$ that is midway between SH0ES and Planck and is practically in no-tension with any of them. This complicates further the issue. Ref. \cite{Yuan-etal-2019-TRGB-local} redid the TRGB analysis with different calibration methods and found a high value of $H_0$, extending the debate. 
 Ref. \cite{VTR2019} provided a discussion showing a persistent $H_0$ tension when data are grouped into various early-time versus late-time data sets. Meanwhile, a number of proposals for extensions to the $\Lambda$CDM-model that attempt to address this tension have been made in, for example, \cite{Poulin-etal-2019-EDE,Blinov-etal-2019-interacting-v-H0,2019-Vagnozzi-H0tension, 2019arXiv190512618L, 2019arXiv190703778D, 2018PhRvD..97l3504P, 2019PhRvD..99d3519G, 2019ApJ...871..210E, 2019PhRvD.100b3505C, 2019arXiv190200534K, 2017JHEP...09..033P, 2017-Zhao-etal-nature-astronomy, 2017EPJC...77..418F, 2017PhRvD..96h3513D, 2017PhRvD..96d3503D, 2017PhLB..774..317S, 2019PhRvD..99j3526K, 2018-Bolejko.K, 2016-Bernal-Verde-Riess-H0, 2016-Valentino-Melchiorri-Silk-reconciling-Planck-local}.

Next comes the tension in the amplitude of matter fluctuations, $\sigma_8$, when comparing the results from (LSS), such as weak gravitational lensing and galaxy clustering, against CMB measurements from WMAP or Planck, see for example \cite{2019-Wibking,2019-Joudaki-Kids+DES,2011WiggleZ-growth-rate,2019-Lange-etal,2016Bernal-etal-param-splitting,2014sdssIII-redshift-space,2015Planck-SZ-cluster-count,2009-Chandra-clustering}. The amplitude is often found to be smaller when measured from LSS surveys. Similar to the case of tension in $H_0$, some attempts have been made to remedy this by considering extensions to the $\Lambda$CDM-model, see for example \cite{Blinov-etal-2019-interacting-v-H0,2019arXiv190804281D}.

Third, several papers have argued that tensions between data sets should be measured by considering the whole cosmological parameter space rather than 1D or 2D marginalized constraints \cite{WL2017a,2017-KiDS-Weak-lensing,WL2017b,Handley-etal-2019-suspiciousness}. A theorem was provided in \cite{WL2017a} to show show that marginalization can only reduce inconsistencies when present in the whole parameter space. This point was also discussed most recently in Ref. \cite{Handley-etal-2019-suspiciousness}. In the context of comparing entire parameter spaces, Ref. \cite{WL2017a} showed that a moderate to strong tension exists between the expansion (geometry) data and the growth of structure from LSS data. Corroborating results can be found in \cite{2015Ruiz-etal-param-splitting,2016Bernal-etal-param-splitting}.

Tensions and inconsistencies between cosmological parameters as determined from different data sets can be due to systematics in the data or possible problems with the underlying model or theory. 
A number of studies have shown how systematic effects can shift and bias the values of cosmological parameters obtained from the data. For example, Refs. \cite{BK2007,KEB2016,JI2017} showed how intrinsic alignment of galaxies, if not taken into account in the analysis, can bias cosmological parameter values by significant amounts (10\% to 50\%). Similarly, modifications to the standard  $\Lambda$CDM-model models have been shown to affect the values obtained for such parameters. Some pioneering work on dark energy parameter discordance due to problems with the underlying model can be found in Ref. \cite{2006-Ishak-splitting}. It is worth noting that such changes in the underlying model can be due to fundamental changes in the theory such as departures from General Relativity, see e.g. the reviews \cite{2012-Clifton-MG,KOYAMA2016TestGR,2015-rev-Joyce-et-al,2016-Joyce-Lombriser-Schmidt-DEvsMG,IshakMG2019}, or simply changes to other exact solutions of Einstein's equations representing other general relativistic cosmological models, see e.g.  \cite{Stephani:2003tm,Bolejko2011,Ishak2012,Peel2014}. 

An important task is to derive mathematical measures and metrics that can detect such inconsistencies when present between data sets and accurately estimating their significance. This has been the subject of active developments and a number of measures have been proposed in, for example \cite{2006-Marshall-etal-Bayesian,2011Robustness-March-etal,2013-tension-Verde-etal,2015-discordance-MacCrann-etal,2015-Battye-Charnock-Moss-difference-vector,2017-Charnock-Battye-Moss,2014-rel-entropy-Seehars-etal,2016-quantify-concor-Seehars-etal,2016-Grandis-information-Gains,2017MNRAS-Joudaki-etal-DIC-inconsistency,WL2017a,2018-Adhikari-Huterer,2018-Raveri-Hu,Handley-etal-2019-suspiciousness,Park-Rozo-2019,Nicola-Amara-Refregier-2019,Kohlinger-etal-2019} (this is only a partial list) and Ref. \cite{WL2017a,2017-Charnock-Battye-Moss,2018-Raveri-Hu} for a comparison or overview of such discordance measures.  

In this paper, we focus on measures that are based on differences between model parameters. In section \ref{sectionI:stability-KL}, we analyze the stability question of such metrics with or without the Karhunen-Loeve (KL) \cite{Karhunen1947,Loeve1978} decomposition. We then discuss in section \ref{section-dimensionality} the question of interpretation of the measures and the effect of the KL decomposition. We provide illustrative constructions and also use the example of WMAP versus Planck to illustrate some of the points made. Afterwards, in section \ref{section:application-current-data} we apply the Index of Inconsistency metric to analyze the tension in $H_0$ between various data sets as well as compare LSS data versus Planck-2018. Finally, in section \ref{section-conclusions} we provide some final remarks and conclusions.

%%%%%%%%%%%%%%%%%%%%%%%%%%%%%%%%%%%%%%%%%%%%%%%%%%%%%%%%%%%%%%%%%%%%%%%
\section{Stability of inconsistency measures and Karhunen-Loeve decomposition} \label{sectionI:stability-KL}
%%%%%%%%%%%%%%%%%%%%%%%%%%%%%%%%%%%%%%%%%%%%%%%%%%%%%%%%%%%%%%%%%%%
One of the important requirements that a measure of discordance must have is to be numerically stable. Specifically, it should lead to the same results regardless of how many times one derives constraints and does a consistency analysis on the same set of experiments or surveys. In other words, one should get the same level of consistency or inconsistency when repeating the MCMC analysis over the same data sets. Then, the standard deviation associated with the measure of discordance should be small. Therefore in this section we explore the numerical stability of IOI and $Q_{\text{UDM}}$ through illustrative constructions and real data. Moreover, we analyze the numerical stability associated with both measures in the context of the so called Karhunen-Loeve (KL) transform \cite{Karhunen1947,Loeve1978,96-KL-Tegmark}. Additionally, we establish a one-to-one relationship between IOI and $Q_{\text{UDM}}$ in the Gaussian and weak-prior limit.

%%%%%%%%%%%%%%%%%%%%%%%%%%%%%%%%%%%%%%%%%%%%%%%%%%%%%%%
\subsection{Index of inconsistency IOI}

The index of inconsistency (IOI) is a moment-based quantity measuring the inconsistencies between two or more constraints and was originally proposed in \cite{WL2017a} and used in for example \cite{WL2017b,XXL-survey-C1-cluster-counts,2018-Camarena-Marra,2019-Camarena-Marra,2019-Vagnozzi-H0tension}. 
We refer the reader to \cite{WL2017a} for background material, a detailed derivation, and an extensive description. 
In summary, for two constraints, the two-data set IOI is defined as
\begin{equation}
\text{IOI} = \frac{1}{2} \boldsymbol{\delta}_{\text{IOI}}^{T} ( \boldsymbol{C_1} + \boldsymbol{C_2} )^{-1} \boldsymbol{\delta}_{\text{IOI}},
\label{IOIeq:two-exp}
\end{equation}
where $\boldsymbol{C_i}$ is the covariance matrix obtained from the posterior associated with the $i$-the data set, while 
\be
\boldsymbol{\delta}_{\text{IOI}}=\boldsymbol{\mu_1}-\boldsymbol{\mu_2}
\ee
is the difference between the mean values of parameters for the two experiments. For multiple constraints, the multi-data set IOI is defined as
\begin{equation}
\text{IOI} = \frac{1}{N} \sum_{i=1}^{N} \left( {\boldsymbol{\mu_i}^T \boldsymbol{L_i}} \boldsymbol{\mu_i} - {\boldsymbol{\mu}}^T \boldsymbol{L} \boldsymbol{\mu} \right),
\label{IOIeq:mult-exp}
\end{equation}
where $\boldsymbol{L_i}=\boldsymbol{C_i}^{-1}$, $\boldsymbol{L}=\sum\boldsymbol{L_i}$, and $\boldsymbol{\mu}=\boldsymbol{L}^{-1}\sum \boldsymbol{L_i \mu_i}$ 
is the Fisher-matrix weighted average mean of all the experiments. The above multi-data set IOI reduces to the quadratic two-data set IOI for two-constraint cases. 

It was shown in \cite{WL2017a,WL2017b,IOI-remarks} that IOI accurately traces inconsistencies when present in a number of illustrative constructions; applications to various current data sets were also demonstrated. An eigen-value decomposition of IOI was derived there as well, and comparisons to a number of inconsistency measures proposed in the literature were provided. A theorem was also given in \cite{WL2017a} showing that marginalization over parameters can only underestimate the level of inconsistency, and therefore confirms the importance of performing consistency analyses in full parameter spaces, as stated in a few other earlier studies \cite{2017-KiDS-Weak-lensing}. This effect of marginalization was also appreciated in a recent study \cite{2019-tensions-WillHandley}. We discuss further properties of IOI and apply it to current data sets in the next sections.

\begin{table}[t]
%\scriptsize
\begin{tabular} { >{\centering}p{3cm} | >{\centering\arraybackslash}p{3.5cm}  >{\centering\arraybackslash}p{3.5cm}  >{\centering\arraybackslash}p{3.5cm}  >{\centering\arraybackslash}p{3.5cm} }
\hline\hline

& & & & \\ 

 Ranges & IOI$<1$  & $1<$IOI$<2.5$ & $2.5<$IOI$<5$ & IOI$>5$   \\ [5pt] \hline

& & & & \\ 

 Interpretation & No significant inconsistency  & Weak inconsistency & Moderate inconsistency & Strong inconsistency   \\ [5pt] \hline

& & & &  \\ 

 Confidence level (one dimension only) & $<1.4$-$\sigma$  & $1.4$-$\sigma - 2.2$-$\sigma$  & $2.2$-$\sigma - 3.2$-$\sigma$ & $>3.2$-$\sigma$   \\ [15pt] 

\hline\hline
\end{tabular}
\caption{The values of IOI can be interpreted using the Jeffreys' scale. We use an overall more conservative terminology than the original one suggested in \cite{jeffreys1998theory}. In the second row we show the comparison between IOI and the commonly used significance level of tension valid only for one-dimensional Gaussian distributions. Note that in \cite{IOI-remarks} it was argued that such a correspondence between Jeffreys' scales interpretation of IOI and n-$\sigma$ can be adopted even for higher dimensional parameter spaces.} 
\label{Table:JeffreyScale} 
\end{table}

%%%%%%%%%%%%%%%%%%%%%%%%%%%%%%%%%%%%%%%%%%%%%%%%%%%%%%%

\subsection{Update difference in mean $Q_{\text{UDM}}$ (applies to two data sets)} \label{section-QUDM}

Sometime after the work on IOI appeared in the literature another measure of discordance based on parameter-differences, known as the update difference in mean ($Q_{\text{UDM}}$), was proposed in \cite{2018-Raveri-Hu}. The idea is that given an initial data set with Gaussian distributions in the parameters, one can update this data set with another experiment and then compare distributions from the first experiment with distributions coming from the joint analysis of the two. 
While the first data set used must be nearly Gaussian in the parameters, there is no such restriction on the second data set. In this way, $Q_{\text{UDM}}$ can measure inconsistencies between two data sets. It was defined in \cite{2018-Raveri-Hu} as
\begin{equation}
Q_{\text{UDM}} = \boldsymbol{\delta}_{\text{UDM}}^{T}( \boldsymbol{C_1} - \boldsymbol{C_{12}} )^{-1} \boldsymbol{\delta}_{\text{UDM}},
\label{QUDM-def}
\end{equation}
where 
\be
\boldsymbol{\delta}_{\text{UDM}}= \boldsymbol{\mu_1}-\boldsymbol{\mu_{12}}
\ee 
is the difference between the mean of the first data set and the mean of the joint data set. 

Furthermore, \cite{2018-Raveri-Hu} applied the Karhunen-Loeve (KL) decomposition to $Q_{\text{UDM}}$ in order to minimize the numerical noise from MCMC sampling.
The value for the update difference in mean after performing a KL decomposition and removing the noisy modes was found to be 
\begin{equation}
Q_{\text{UDM}_\text{KL}} = \sum_{\alpha=1}^{N_{\text{KL}}} \frac{(\boldsymbol{\delta}_{\text{UDM}_\text{KL}}^\alpha)^2}{\lambda^\alpha - 1}.
\label{QUDM-defKL}
\end{equation}
Here, $\boldsymbol{\delta}_{\text{UDM}_\text{KL}}= \boldsymbol{\Phi}^T \boldsymbol{\delta}_{\text{UDM}}$, where $\boldsymbol{\Phi}$ represents the matrix of eigenvectors associated with the modified eigenvalue problem given by
\begin{equation}
\boldsymbol{C_1} \boldsymbol{\Phi} = \boldsymbol{\Lambda}\boldsymbol{C_{12}} \boldsymbol{\Phi},
\label{QUDM-eig0}
\end{equation}
with the conditions 
\begin{equation}
\boldsymbol{\Phi}^T \boldsymbol{C_1} \boldsymbol{\Phi} = \boldsymbol{\Lambda}
\label{QUDM-eig1}
\end{equation}
and
\begin{equation}
\boldsymbol{\Phi}^T \boldsymbol{C_{12}} \boldsymbol{\Phi} = \boldsymbol{I}.
\label{QUDM-eig2}
\end{equation}
The values $\lambda^\alpha$ are the eigenvalues obtained from the eigenvalue matrix $\boldsymbol{\Lambda}=\text{diag}(\lambda_1, \lambda_2, ..., \lambda_N)$ after solving (\ref{QUDM-eig1}).

The dimension of the covariance matrices is the number of parameter $N_{\rm{p}}$. After applying the KL transform the matrix $\Lambda$ becomes diagonal and eigenmodes are decorrelated from each other. The eigenvalues play the role of the variances associated with each eigeinmode. It is suggested in Ref. \cite{2018-Raveri-Hu} that one can remove the numerical noise dominated modes by dropping out the modes with $\lambda^\alpha \simeq 1$. These are the modes whose uncertainty get only slightly smaller when adding the second data set to the first in a joint analysis, and are defined as ``noise-dominated modes'' in Ref. \cite{2018-Raveri-Hu}. If left there, these modes may create large errors in $Q_{\text{UDM}}$, as we explain in section \ref{section-stability}. After removing all the noise-dominated modes, the new dimensionality obtained after this process is labeled as $N_{\text{KL}}$, and it is necessarily that $N_{\text{KL}}\leq N_{\rm{p}}$. Thus we can interpret Eq. (\ref{QUDM-defKL}) as Eq. (\ref{QUDM-def}) after removing the noise-dominated modes in a principal component analysis after a KL decomposition.

%In the case of the update difference in mean, the authors chose to interpret the values

\subsection{Gaussian and weak prior limit}
\label{section:gaussianweaklimit}

\begin{figure}[t]
\centering
\includegraphics[width=8cm]{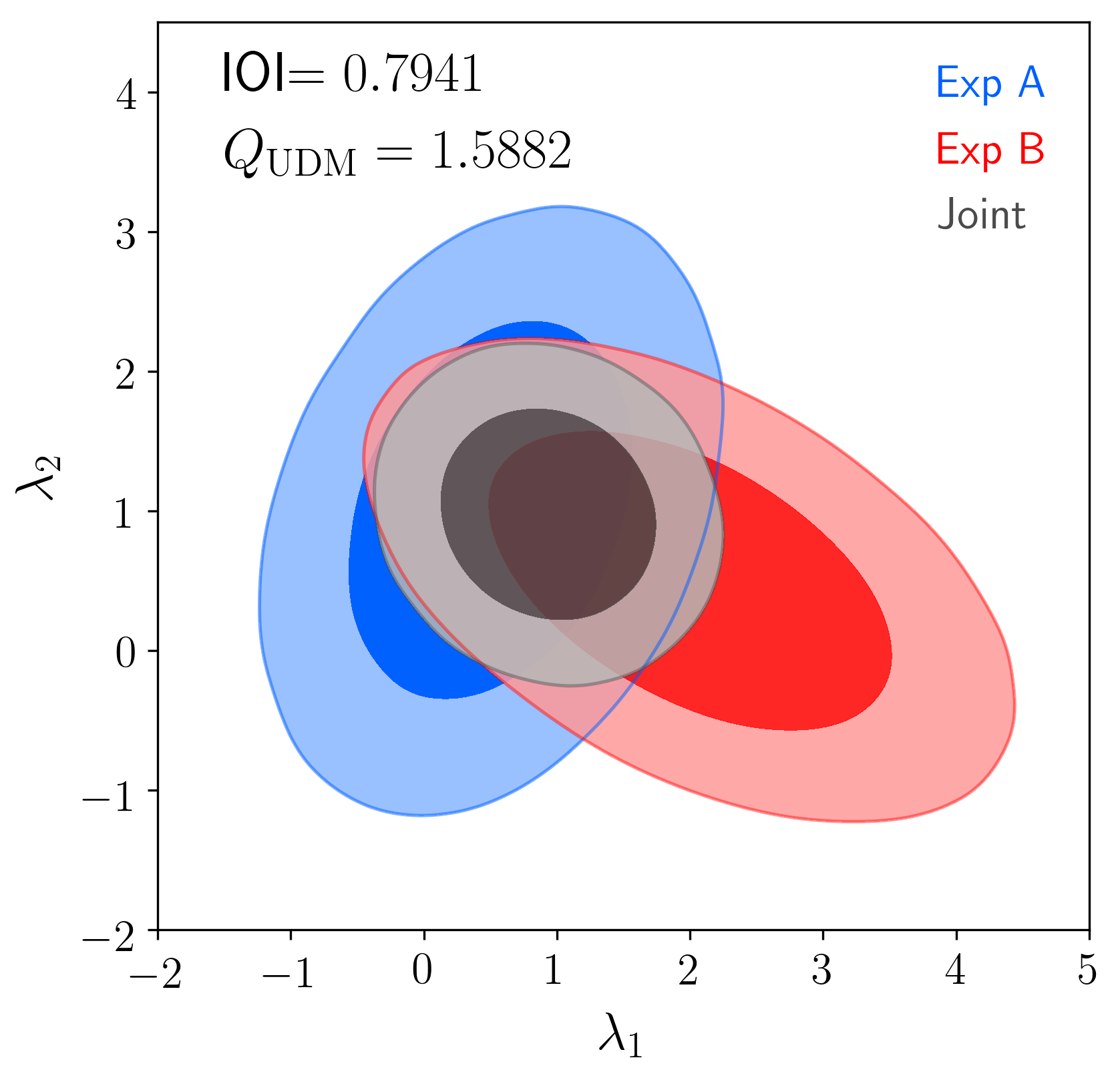}
\caption{Numerical example to illustrate the relation $\text{IOI}=Q_{\text{UDM}}/2$ in (nearly) Gaussian and weak prior limits. We compare two Gaussian distributions and obtain the joint distribution. For the two toy-experiments Exp A and Exp B, the parameter means and covariance matrices are given in sub-section \ref{section:gaussianweaklimit} . We also show in gray the plot for the joint Gaussian analysis. We calculate IOI and $Q_{\text{UDM}}$ by using \eqref{IOIeq:two-exp} and \eqref{QUDM-def}, respectively.}
\label{Fig:Gaussian-weak-prior-limit}
\end{figure}

It is worth mentioning that for Gaussian parameter distributions and in the weak prior limit there is a simple relationship between the two inconsistency measures discussed above that is given by 
\begin{equation}
Q_{\text{UDM}}=2\,\text{IOI}.
\label{IOI-QUDM-relationship}
\end{equation}
We provide here a numerical demonstration of this relationship in Fig. \ref{Fig:Gaussian-weak-prior-limit} as well as an exact analytical proof  detailed in appendix \ref{Appendix:2IOI=QUDM}. 

As we can observe in Fig. \ref{Fig:Gaussian-weak-prior-limit}, after building two Gaussian distributions and obtaining the joint distribution, we can compute IOI for Exp A and Exp B, and compute $Q_{\text{UDM}}$ by using Exp A and Joint. The Gaussian and weak prior limit refers to a situation where likelihoods are Gaussian on the parameters and the prior is weak. 

To produce  Fig. \ref{Fig:Gaussian-weak-prior-limit} and calculate the inconsistency measures, we proceed as follows. 
For the first toy-experiment (Exp A) we use 
\be
\boldsymbol{\mu_A}=\left( \begin{matrix} 0.5 \\ 1 \end{matrix} \right)
\ee
and 
\be
\boldsymbol{C_A}=\left( \begin{matrix} 0.5 & 0.18973 \\ 0.18973 & 0.8 \end{matrix} \right),
\ee
while for the second toy-experiment (Exp B), we set 
\be
\boldsymbol{\mu_B}=\left( \begin{matrix} 2 \\ 0.5 \end{matrix} \right)
\ee
and 
\be
\boldsymbol{C_B}=\left( \begin{matrix} 1 & -0.35355 \\ -0.35355 & 0.5 \end{matrix} \right).
\ee
The ellipses corresponding to the individual experiments are plotted, as well as that of the joint Gaussian analysis (gray). Then we calculate IOI and $Q_{\text{UDM}}$ as given by the respective equations (\ref{IOIeq:two-exp}) and (\ref{QUDM-def}).  The values obtained confirms the relationship (\ref{IOI-QUDM-relationship}) and the analytical proof of appendix \ref{Appendix:2IOI=QUDM}.

Finally, it is worth noting that IOI is a moment-based measure and requires distributions of model parameters to be nearly Gaussian. Therefore, when we want to perform a consistency analysis on two cosmological experiments whose parameter distributions are highly non-Gaussians, we need to combine them separately with other data in order to break degeneracies between parameters and to obtain Gaussian or nearly Gaussian distributions for the parameters.
It is also worth pointing out that an advantage of IOI is that it can be applied to multiple experiments and data sets, and the multi-data set IOI has been applied to various multi-constraint comparisons in Ref. \cite{WL2017b}. Discussion of multiple experiments consistency comparisons  is generally a subject not well covered in literature.

The $Q_{\text{UDM}}$ measure works differently and can be applied to only two data sets at once. However,Ref. \cite{2018-Raveri-Hu} introduced $Q_{\text{UDM}}$ as a measure that requires only one data set to provide a Gaussian parameter distribution, with no requirement of Gaussianty for the second data set, and that along with a KL decomposition it is numerically stable against non-Gaussianity. We will investigate the stability of both IOI and $Q_{\text{UDM}}$ as well as the role of KL decomposition in section \ref{section-stability}. 

Nonetheless, it is fair to assert that there are currently no \textit{perfect} discordance measures that can be simultaneously applied to multiple data sets, cover non-Gaussian cases and accurately trace inconsistencies when they are present \cite{WL2017a,2011Robustness-March-etal,2013-tension-Verde-etal,2015-discordance-MacCrann-etal,2016-quantify-concor-Seehars-etal,2016-tensions-Grandis-etal,2017MNRAS-Joudaki-etal-DIC-inconsistency,2018-Raveri-Hu,2018-Adhikari-Huterer,Handley-etal-2019-suspiciousness}.

\subsection{Stability analysis} \label{section-stability}
\begin{table}[t]
\begin{ruledtabular}
\begin{tabular} { c | c  c  c|  c  c  c }

\multirow{5}{*}{Experiment 1 .vs. Experiment 2} & \multicolumn{3}{c|}{ } & \multicolumn{2}{c}{ }  \\ 
 & \multicolumn{3}{c|}{ $R-1 < 0.01 $} & \multicolumn{3}{c}{$R-1 < 0.05$} \\ [5pt] \cline{2-7} 

 & & & & & & \\ 

 & IOI & $Q_{\text{UDM}}$  & $Q_{\text{UDM}_\text{KL}}$ & IOI & $Q_{\text{UDM}}$  & $Q_{\text{UDM}_\text{KL}}$ \\ [5pt] \hline\hline

& & & & & & \\ 

TTTEEE + CMB lens $\#1$ .vs. DES + Added Probes $\#1$ & $4.11$ & $5.78$ & $5.44$ & $4.08$ & $8.63$ & $4.52$ \\

& & & & & & \\ 

TTTEEE + CMB lens $\#1$ .vs. DES + Added Probes $\#2$ & $3.91$ & $4.66$ & $5.41$ & $3.98$ & $5.68$ & $4.51$ \\

& & & & & & \\ 
 
TTTEEE + CMB lens $\#1$ .vs. DES + Added Probes $\#3$ & $4.05$ & $5.73$ & $5.57$ & $3.86$ & $3.87$ & $4.33$ \\

& & & & & & \\ 
 
TTTEEE + CMB lens $\#2$ .vs. DES + Added Probes $\#1$ & $4.12$  & $6.26$  & $5.49$ & $4.16$ & $5.89$ & $4.89$ \\

& & & & & & \\ 

TTTEEE + CMB lens $\#2$ .vs. DES + Added Probes $\#2$ & $3.92$ & $-4.20$  & $5.45$ & $4.06$ & $4.30$ & $4.82$ \\

& & & & & & \\ 

TTTEEE + CMB lens $\#2$ .vs. DES + Added Probes $\#3$  & $4.06$ & $5.34$  & $5.63$ & $3.93$ & $4.69$ & $4.58$ \\

& & & & & & \\ 

TTTEEE + CMB lens $\#3$ .vs. DES + Added Probes $\#1$ & $4.07$  & $5.55$  & $5.04$ & $4.18$ & $17.19$ & $4.82$ \\

& & & & & & \\ 

TTTEEE + CMB lens $\#3$ .vs. DES + Added Probes $\#2$ & $3.87$ & $4.48$ & $4.99$ & $4.08$ & $-4.72$ & $4.76$ \\

& & & & & & \\ 

TTTEEE + CMB lens $\#3$ .vs. DES + Added Probes $\#3$ & $4.01$ & $6.20$ & $5.13$ & $3.95$ & $6.08$ & $4.52$ \\

& & & & & & \\

\end{tabular}
\end{ruledtabular}
\caption{We compare TTTEEE + CMB lens against DES + Added Probes after obtaining independently the constraints for each of them, a total of three times. We label the three constraints for each experiment with $\#1$, $\#2$ and $\#3$. We also obtain the constraints for two cases: In the first case we set a convergence value of $R-1=0.01$, while in the second case we set $R-1=0.05$. A lower value of $R-1$ produces a higher accuracy in the confidence limits. We calculate IOI, $Q_{\text{UDM}}$ and $Q_{\text{UDM}_\text{KL}}$ in each case, in order to see whether the measure of discordance is stable or not.}
\label{Table:comparison-stability}
\end{table}

\begin{table}[t]
\scriptsize
\begin{tabular} { >{\centering}p{2.8cm} | >{\centering}p{2.8cm} | >{\centering}p{2.8cm} | >{\centering}p{2.8cm} | >{\centering}p{2.8cm} | >{\centering\arraybackslash}p{2.8cm} }
\hline\hline
 \multicolumn{3}{c|}{ }  \\ 
 \multicolumn{3}{c|}{ $R-1 < 0.01 $} & \multicolumn{3}{c}{$R-1 < 0.05$} \\ [5pt] \cline{1-6} 

& & & & & \\ 

 $\text{IOI}$ & $Q_{\text{UDM}}$  & $Q_{\text{UDM}_\text{KL}}$ & $\text{IOI}$ & $Q_{\text{UDM}}$  & $Q_{\text{UDM}_\text{KL}}$ \\ [5pt] \hline

& & & & & \\ 

 $2.14\%$ & $11.08\%$ & $4.14\%$ & $2.52\%$ & $57.85\%$ & $3.85\%$ \\

& & & & & \\ 

\hline\hline
\end{tabular}
\caption{We list $\sigma/\mu \times 100\%$  associated with IOI, $Q_{\text{UDM}}$ and $Q_{\text{UDM}_\text{KL}}$. We find that IOI is a stable measure using two different convergence levels. Furthermore, we find that in each case, the interpretation of IOI is always the same using Jeffrey's scale. On the other hand, $Q_{\text{UDM}}$ seems very unstable giving very high or negative values in some cases. Additionally, we can see that using the KL decomposition helps $Q_{\text{UDM}}$ and makes it stable. However, among the three measures of discordance, IOI seems to be more stable.}
\label{Table:standard-deviation}
\end{table}

We run cosmological parameter estimation and inconsistency measure calculations (for IOI and $Q_{\text{UDM}}$) using the same data sets multiple times. We analyze the values obtained and report standard deviations in each case as we discuss below. We use the $\Lambda$CDM model with six cosmological parameters: $\Omega_bh^2$ and $\Omega_c h^2$, the baryon and cold dark matter physical density parameters, respectively; $\theta$, the ratio of the sound horizon to the angular diameter distance of the surface of last scattering multiplied by $100$; $\tau$, the reionization optical depth; $n_s$, the scalar-perturbation spectral index; and $\ln(10^{10} A_s)$, the amplitude of the primordial power spectrum. In each MCMC sample for this section, we fix the reionization optical depth $\tau$ to 0.06 since it is not constrained by the other probes used. We compare Planck-2018 data against large-scale structure data from the Dark Energy Survey (DES) in combination with some added probes by calculating IOI, Q$_{\text{UDM}}$ and Q$_{\text{UDM}_\text{KL}}$. Specifically, we use the Planck CMB lensing measurements in combination with TT, TE, TT Planck high-$\ell$ temperature likelihood, used in \cite{2018-Planck-cosmo-params} and described in \cite{Planck18-CMBlens}. We label this combination as TTTEEE + CMB lens. On the other side, we use the clustering and lensing data from the DES Year 1 analysis \cite{2017-DESC-1st-joint} in combination with some added probes. We use BAO measurements from the BOSS Data Release 12 \cite{AlamEtAl2016}, together with two BAO data sets from the 6DF Galaxy Survey \cite{2011BAO-6df} and the SDSS Data Release 7 Main Galaxy Sample \cite{2015BAO-sdss-mgs}. We combine the BAO measurements with the supernova Pantheon compilation \cite{Pan-STARRS-2017} given by the combination of 279 Type Ia supernovae (SNe Ia) $(0.03< z <0.68)$ with useful distance estimates of SNe Ia from SDSS, SNLS, various low-z and HST samples, giving a total of 1048 SNe Ia ranging from $(0.01< z <2.3)$. Additionally, we use BBN measurements from \cite{2017Cooke} in order to constrain $\Omega_b h^2$, and we refer to this combination as DES + Added Probes. Finally, we obtain the constraints by using the publicly available Markov Chain Monte Carlo (MCMC) code \texttt{CosmoMC} \cite{COSMOMC}.

We proceed as follows: we compare TTTEEE + CMB lens data against DES + AddedProbes. We run each data set a total of three times, so we can compare the two data sets a total of nine times. We label the first, second and third run of each experiment with $\#1$, $\#2$ and $\#3$, respectively. For a fixed convergence limit, a measure with a smaller standard deviation is more numerically stable. Furthermore, we do this process for two different chain convergence values, using a simple generalized Gelman-Rubin statistic $R-1$ \cite{2013-Lewis-EfficientSampling,Gelman-Rubin}, with  values of $0.01$ and $0.05$. Here, a value of $R-1=0.01$ is numerically more accurate than $R-1=0.05$. Therefore, we expect a lower standard deviation for IOI and $Q_{\text{UDM}}$ when $R-1$ is set to a lower value. Once we obtain the cosmological constraints, we compute IOI using equation (\ref{IOIeq:two-exp}), $Q_{\text{UDM}}$ given by (\ref{QUDM-def}) and $Q_{\text{UDM}_\text{KL}}$ using (\ref{QUDM-defKL}). The results for each measure are compiled in Table \ref{Table:comparison-stability}\footnote{In this case, the relationship $Q_{\text{UDM}}=2\text{IOI}$ does not hold. Possible reasons for this might be the non-Gaussianity of the parameter distributions. We can also see that $Q_{\text{UDM}_\text{KL}}\neq2\text{IOI}$, this could be due to the removal of some noise-dominated modes in addition to the non-Gaussiantiy. Indeed, if we use distributions that are closer to Gaussian ones, such as the constraints from Planck and WMAP, the relation $Q_{\text{UDM}}=2\text{IOI}$ holds very well; see section \ref  {section:WMAP-Planck}. Also $Q_{\text{UDM}_\text{KL}}=2\text{IOI}$ holds very well too, since only one KL mode is removed.} and the associated standard deviations divided by the mean values in Table \ref{Table:standard-deviation}.

As we can see in Table \ref{Table:comparison-stability}, IOI shows very similar values for each analysis. One can observe that this is consistently the case for the two convergence levels used. On the other hand, $Q_{\text{UDM}}$ shows a much wider range of values with some of them being too large (e.g. $17.19$ for $\#3$ vs $\#1$), for the case of $R-1 \leq 0.05$. We also obtain negative values (e.g. $-4.72$ for $\#2$ vs $\#2$), which highlights some stability issues if used without KL decomposition. This instability in $Q_{\text{UDM}}$ can be seen from equation (\ref{QUDM-def}). For Gaussian cases with no numerical noises, the factor $( \boldsymbol{C_1} - \boldsymbol{C_{12}} )^{-1}$ is positive definite. But due to non-Gaussianity or just numerical noise, it is not the case anymore. The fact that $( \boldsymbol{C_1} - \boldsymbol{C_{12}} )^{-1}$ is not positive definite in practice gives the possibilities of getting negative values of $Q_{\text{UDM}}$, which are difficult to interpret. This can be seen more clearly in KL modes; see our latter discussion. After the KL decomposition and filtering of modes associated with numerical noise are performed, $Q_{\text{UDM}_\text{KL}}$ becomes numerically stable. The KL decomposition decorrelates the modes and one then sorts the eigen-modes by order of improvement (i.e., by order of descending $\lambda^\alpha$). If the combined constraints only slightly improve the uncertainty on a particular KL mode compared to the first constraint, the corresponding $\lambda^\alpha$ will be closer to $1$. The corresponding KL modes are numerically noisy and must be removed according to some threshold criterion. After removing the noisy modes (five modes for this case), one obtains $Q_{\text{UDM}_\text{KL}}$ with $N_{\text{KL}}=1$ and it then becomes stable.  

We express this dispersion by showing in Table \ref{Table:standard-deviation} the standard deviation divided by the mean value associated with each measure of discordance. However, in the case of $Q_{\text{UDM}}$ we do not consider the negative values, since they cannot be used to distinguish if an inconsistency is present. As shown there, $Q_{\text{UDM}}$ has a large dispersion of $11.08\%$ and $57.85\%$. After removing the noisy modes using the KL decomposition process, the standard deviations for $Q_{\text{UDM}_\text{KL}}$  are reduced to more acceptable values (i.e. $4.14\%$ and $3.85\%$) but they still remain more than twice the small values reached by IOI without any KL decomposition (i.e. $2.14\%$ and $2.52\%$). 

In other words, we find that IOI has very stable values with very little dispersion. This is contrary to what was claimed in \cite{2018-Raveri-Hu} about measures of this type without any supporting specifications.

Finally, we show explicitly in what follows that the stability of IOI does not depend on the KL decomposition.  
Indeed, we can decorrelate the covariance matrices involved in the calculation of IOI (see Eq. (\ref{IOIeq:two-exp})) by using an analogous modified eigenvalue problem to the one shown for $Q_{\text{UDM}}$ in section \ref{section-QUDM}. We consider the modified eigenvalue problem as
\begin{equation}
\boldsymbol{C_1} \boldsymbol{\Phi} = \boldsymbol{\Lambda}\boldsymbol{C_2} \boldsymbol{\Phi},
\label{IOI-eig0}
\end{equation}
where the covariance matrices are subject to the conditions
\begin{equation}
\boldsymbol{\Phi}^T \boldsymbol{C_1} \boldsymbol{\Phi} = \boldsymbol{\Lambda}
\label{IOI-eig1}
\end{equation}
and
\begin{equation}
\boldsymbol{\Phi}^T \boldsymbol{C_2} \boldsymbol{\Phi} = \boldsymbol{I}.
\label{IOI-eig2}
\end{equation}
This leads to the following decorrelated form of IOI 
\begin{equation}
\text{IOI} = \sum_{\alpha} \frac{(\boldsymbol{\delta}_{\text{IOI}_\text{KL}}^\alpha)^2}{\lambda^\alpha + 1},
\label{IOI-KL}
\end{equation}
where $\lambda^\alpha$ are the elements of the diagonal matrix $\Lambda$ and $\boldsymbol{\delta}_{\text{IOI}_\text{KL}} = \boldsymbol{\Phi}^T \boldsymbol{\delta}_{\text{IOI}}$. In KL modes, both $\boldsymbol{C_1}$ and $\boldsymbol{C_{12}}$ are diagonalized. 

A quick comparison between expressions (\ref{QUDM-defKL}) and (\ref{IOI-KL}) can tell us why IOI is stable by itself, while $Q_{\text{UDM}}$ needs a removal process for noisy modes. In the case of the update difference in mean, we can have problems in two cases. The first case is that the factor $( \boldsymbol{C_1} - \boldsymbol{C_{12}} )$ is not always definite positive after MCMC sampling. This is because the eigenvalues $\lambda^\alpha$ can be smaller than $1$, allowing for the possibility of a negative $Q_{\text{UDM}}$ as seen in Table \ref{Table:comparison-stability}. The second case is when the first experiment has more constraining power than the second experiment in some particular modes. In this case the second experiment does not improve the constraint on a given mode much over the first one and this leads to $\lambda^\alpha\approx 1$. These modes can be identified after a KL transform and removed. Here, we remove the modes that satisfy the condition $\lambda-1<1/\lambda_{max}$, where $\lambda_{max}$ is the biggest mode. Note that this includes all modes with $\lambda^\alpha < 1$.

Now, in the case of IOI, two very similar experiments would lead to $\lambda^\alpha\approx 1$, but this would not drastically increase IOI. Moreover, if experiment 1 has more constraining power than experiment 2, this would lead to eigenmodes with $\lambda^\alpha  \rightarrow  0$. On the other hand, if experiment 2 is more constraining than experiment 1, we would observe some eigenmodes with $\lambda^\alpha  \rightarrow  \infty$. In any case, modes with $\lambda^\alpha  \rightarrow  0$ would not affect (\ref{IOI-KL}) and $\lambda^\alpha  \rightarrow  \infty$ would have negligible contributions. Finally, it is worth noting that the factor $( \boldsymbol{C_1} + \boldsymbol{C_2} )$ involved into the computation of IOI is always positive definite. Thus, the stability of IOI found numerically further above is confirmed analytically here.

%%%%%%%%%%%%%%%%%%%%%%%%%%%%%%%%%%%%%%%%%%%%%%%%%%%%%%%
%%%%%%%%%%%%%%%%%%%%%%%%%%%%%%%%%%%%%%%%%%%%%%%%%%%%%%%
%%%%%%%%%%%%%%%%%%%%%%%%%%%%%%%%%%%%%%%%%%%%%%%%%%%%%%%%%%%%%%%%%%%%%%%
\section{Parameter spaces, interpretation and KL decomposition} \label{section-dimensionality}
%%%%%%%%%%%%%%%%%%%%%%%%%%%%%%%%%%%%%%%%%%%%%%%%%%%%%%%%%%%%%%%%%%%%%

The interpretation of the measures of discordance is crucial in order to accurately represent the degree of inconsistency between cosmological data sets. Different approaches to interpret these measures have been discussed in the literature \cite{2019-tensions-WillHandley,2017-Charnock-Battye-Moss,2015-Battye-Charnock-Moss-difference-vector,2014-rel-entropy-Seehars-etal,2006-Marshall-etal-Bayesian,2015MNRAS.449.2405K,2013-tension-Verde-etal}. A recent detailed discussion of the interpretation can be found in \cite{IOI-remarks} while we focus here on analyzing the question with respect to the decomposition of measures.

%%%%%%%%%%%%%%%%%%%%%%%%%%%%%%%%%%%%%%%%%%%%%%%%%%%%%%%
 
\subsection{Interpretation of measures and KL mode-filtering algorithms}
\label{section:KLmodes}
%%%%%%%%%%%%%%%%%%%%%%%%%%%%%%%%%%%%%%%%%%%%%%%%%%%%%%%

\begin{figure}[t]
\begin{center}
 {\includegraphics[width=8cm]{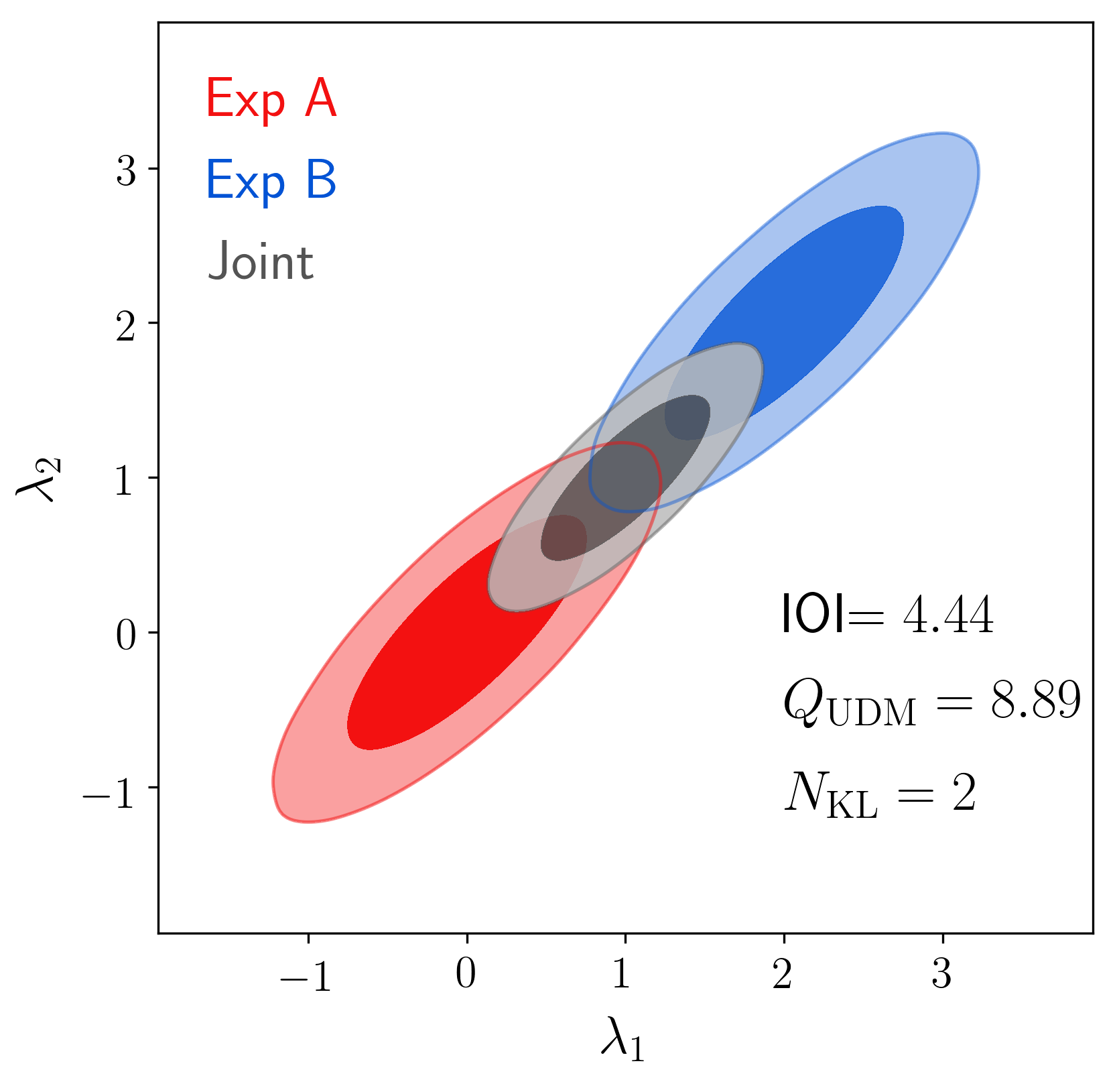}} 
\end{center}
\caption{Numerical example that shows that there are cases where the number of degrees of freedom will not change after the KL decomposition. This happens when the second data set provides a considerable improvement in each mode over the first data set (Gaussian). The parameter means and covariance matrices are given in sub-section \ref{section:KLmodes}. For this plot $N=N_{\text{KL}=}2$ so there is no reduction of the degrees of freedom. 
One can see easily that such a construction can be generalized to  $N$ parameters and have $N_{\text{KL}}=N$. See section \ref{section:KLmodes} for further discussion and implication for the interpretation of the degree of inconsistency.}
\label{Fig:toymodel-NKL}
\end{figure}

\begin{figure}[h!]
\begin{center}
 {\includegraphics[width=8cm]{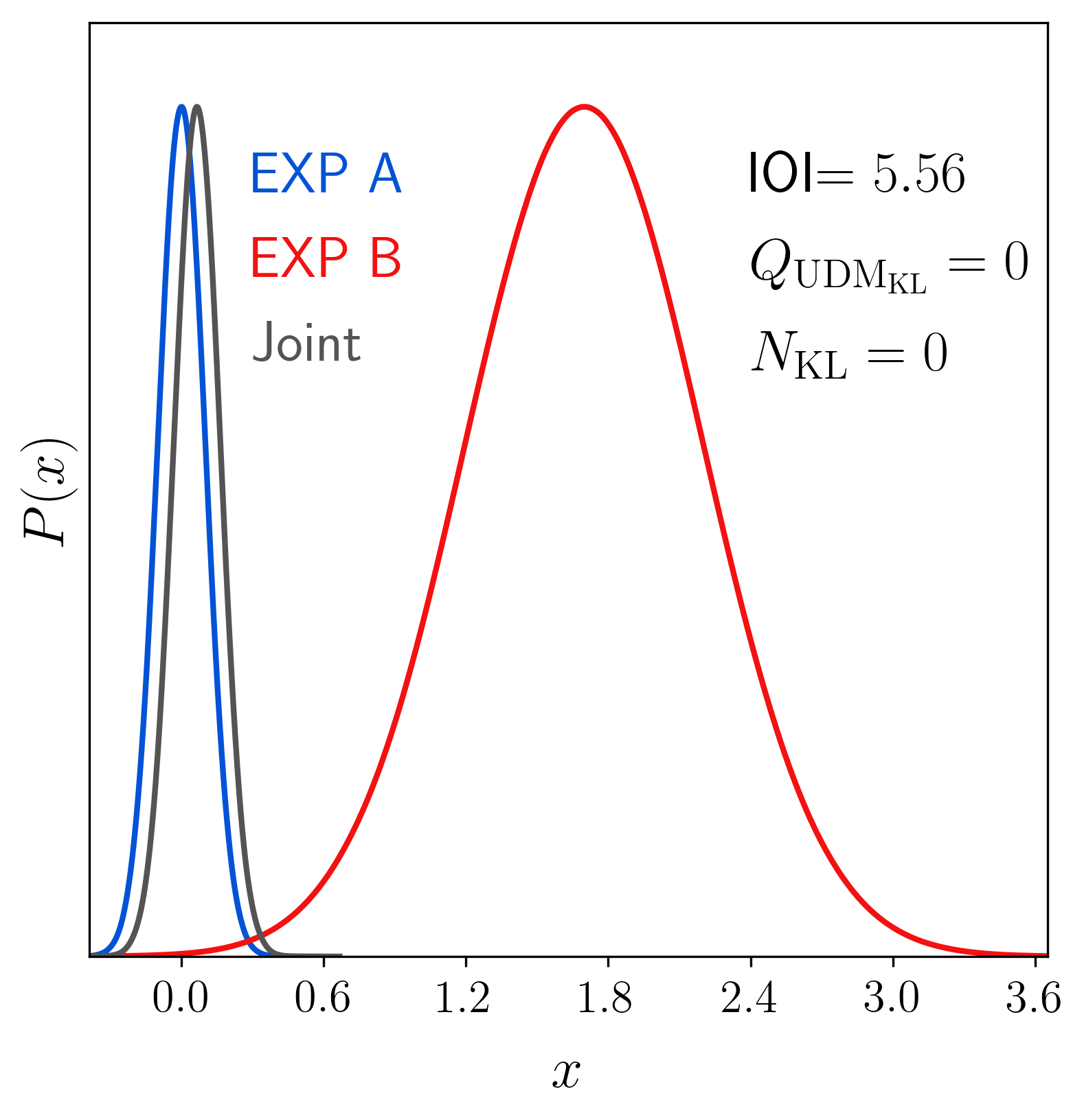}} 
\end{center}
\caption{One dimensional constraints for an illustrative construction model where we can clearly observe that, although the algorithm for removing noise dominated modes used to compute $Q_{\text{UDM}_\text{KL}}$ provides some numerical stability to the measure, it is not designed to deal with some specific situations where a strong inconsistency might be present. We also show the value of IOI which is numerically stable and also can properly trace this inconsistency. Here we use $\mu_A=0$, $\mu_B=1.7$, $\sigma_A=0.1$ and $\sigma_B=0.5$. The interpretation of $Q_{\text{UDM}_\text{KL}}=0$ and $N_{\text{KL}}=0$ is discussed in \cite{2018-Raveri-Hu}.}
\label{Fig:algorithm-NKL}
\end{figure}

The interpretation of the values of IOI is done via Jeffreys' scales. Jeffreys's scales were empirical scales originally suggested for the interpretation of the Bayesian evidence ratio \cite{jeffreys1998theory}. It was suggested in  \cite{WL2017a} to use such scales for IOI because they matched well with the inconsistencies in the illustrative cases studied there and also some other measures that reduce or relate to IOI in Gaussian cases had used these Jeffreys's scales \cite{WL2017a}. In that way, it was found practical to use the same scales to be able to compare these measures to one another. However, it worth noting that we use an overall more conservative terminology than in the original Jeffreys's scales paper \cite{jeffreys1998theory} and similar to the one used in, for example, Ref. \cite{Trotta2008-model-select} as shown in Table \ref{Table:JeffreyScale}. Nonetheless, we point out that the category ``moderate'' in Table \ref{Table:JeffreyScale} does not mean ``insignificant'' or ``ignorable''. On the contrary, one should understand ``moderate'' here as an indicator of a ``non-ignorable'' inconsistency. Finally, we provide a one-to-one relation between IOI and the commonly used confidence level of tension using the relation $n \text{-} \sigma = \sqrt{2\text{IOI}}$. One can observe then some agreement between Jeffrey's scale and the n-$\sigma$ language and the word ``moderate'' is, in fact, an indication of a non-ignorable inconsistency and in particular when IOI is in the high end of this range. The problem of significance and interpretation of inconsistency measures is a subject of ongoing debate in the literature and we refer the reader to discussions in for example Ref. \cite{IOI-remarks}. 

It was argued in \cite{IOI-remarks} that one should not use the conversion from IOI or a function of it to a probability to exceed (PTE) that depends on the number of parameters considered as for example done for $Q_{\text{UDM}}$.
It was shown there that using higher parameter degrees of freedom in such a conversion results in underestimating inconsistencies.

Before we comment on the KL decomposition, we provide here an example to illustrate the argument presented in \cite{IOI-remarks}. Let us consider a model with one parameter $\lambda_1$ and let us assume that one experiment measures $\mu_1=0$ and $\sigma_1^2=0.5$, while a second experiment measures $\mu_2=\sqrt{8}$ and $\sigma_2^2=0.5$. If we compute IOI using (\ref{IOIeq:two-exp}) we obtain IOI$=4$, this indicates a strong inconsistency that may be caused because of systematics in one of the experiments (or both), or problems with the underlying model. Another way to interpret this value is consider that IOI follows a $\chi$-square distribution and compute $\text{PTE}=\frac{\Gamma\left( \frac{1}{2}N_{\text{KL}},\text{ IOI} \right)}{\Gamma\left(\frac{1}{2}N_{\text{KL}}\right)}$, where $\Gamma(\cdot)$ represents the gamma function, while $\Gamma(\cdot; \cdot)$ is the upper incomplete gamma function. For cases with only one parameter as here, the interpretation provided by this approach is suitable and is consistent with that of Jeffreys' scales of IOI. Indeed, it will give us that $\text{PTE}=0.46\%$ which indicates a significant inconsistency. However, let us extent the model to a 10-parameters model where the first parameter follows the same distribution as before for both of the experiments, but the other 9 parameters show exactly the same distribution when they are measured by any of the two experiments. More specifically, the first experiment now measures $\boldsymbol{\mu_1}=(0,0,...,0)$ and $\boldsymbol{C_1}=\text{diag}(0.5,0.5,...,0.5)$ (diag means diagonal matrix), while the second experiment measures $\boldsymbol{\mu_2}=(\sqrt{8},0,...,0)$ and $\boldsymbol{C_2}=\text{diag}(0.5,0.5,...,0.5)$. If this is the case, we would still obtain $\text{IOI}=4$ which according to the Jeffrey's scale indicates a strong inconsistency, which is still given by the first parameter. However, since the number of degrees of freedom was increased by the addition of extra parameters, we obtain $\text{PTE}=62.88\%$ and the inconsistency is not significant anymore. Therefore, this example corroborates with  \cite{IOI-remarks} and confirms that even in a case where we know that a given parameter shows a severe inconsistency, interpreting IOI using a PTE and significance level as describe above may lead to underestimate the inconsistency for a high dimensional parameter space.

Now, even though in many practical cases, the KL number of degrees of freedom $N_{\text{KL}}$ appears to be small (1, 2 or even 0) for $Q_{\text{UDM}}$ \cite{2018-Raveri-Hu}, we show using illustrative constructions and a real data example that there exists cases where $N_{\text{KL}}$ can remain large. A high value of $N_{\text{KL}}$ may be explained by a significant improvement of constraint in each mode. But as we explained above, such a large $N_{\text{KL}}$ can lead to underestimate an inconsistency when converted in terms of PTE and significance level. Right below are two illustrative numerical examples and a concrete example using real data from WMAP versus Planck is given in the next sub-section. In the three examples, this conversion leads to underestimation of  inconsistencies. 

We show in Fig. \ref{Fig:toymodel-NKL} an illustrative example where we can have $N_{\text{KL}}$ as large  as the original number of parameters with no reduction. Specifically, we consider a model given by the parameters $\lambda_1$ and $\lambda_2$, which follow Gaussian distributions for both experiment A (Exp A) and experiment B (Exp B) as follows. For EXP A, the mean value is given by
\be
\boldsymbol{\mu_A}=\left( \begin{matrix} 0 \\ 0 \end{matrix} \right),
\ee
while the associated covariance matrix is
\be
\boldsymbol{C_A}=\left( \begin{matrix} 0.25 & 0.2 \\ 0.2 & 0.25 \end{matrix} \right).
\ee
Similarly, the mean value of EXP B is
\be
\boldsymbol{\mu_B}=\left( \begin{matrix} 2 \\ 2 \end{matrix} \right)
\ee
and the covariance matrix of EXP B is the same that $\boldsymbol{C_A}$, then
\be
\boldsymbol{C_B}=\left( \begin{matrix} 0.25 & 0.2 \\ 0.2 & 0.25 \end{matrix} \right).
\ee
Using Exp A and Exp B we can calculate the value of IOI to be 4.44. Additionally, we can obtain the joint distribution and compute $Q_{\text{UDM}}=8.89$, but even when we use the KL decomposition, Exp B offers a significant improvement on both modes so the algorithm for filtering modes after the KL decomposition does reduce  $N_{\text{KL}}$. An exact similar example in which $N_{\text{KL}}$ does not drop can be constructed in a higher parameter space. For such cases, the large  $N_{\text{KL}}$  will still affect the conversion to PTE which will underestimate the inconsistency. 

Finally, it is important to mention that the KL decomposition process allows to analyze a measure of discordance in a more convenient basis. However, removing the numerical noise from a measure ultimately depends on the algorithm used to removes the noisy modes. The authors in \cite{2018-Raveri-Hu} presented an algorithm that allows to identify the numerical noise dominated modes. However, as they pointed out, if we combine a data set that is very constraining with a data set that weakly constrains the parameters, the algorithm used will drop all the modes, even when an inconsistency may be present. In Fig. \ref{Fig:algorithm-NKL} we show an illustrative example where two experiments present a strong inconsistency. \textit{Nevertheless, we can see that the algorithm for filtering modes after KL decomposition presented in \cite{2018-Raveri-Hu} will drop the $x$ parameter from the analysis, since its improvement is not big enough, even when an inconsistency was originally present}. 

On the other hand, we show in Fig. \ref{Fig:algorithm-NKL} that IOI can deal with such cases without requiring any further algorithm for filtering modes. This example shows a case where clearly IOI can properly trace an inconsistency with no need to discard valuable information via filtering mode algorithms. As we showed in section \ref{section-stability} since, IOI is a stable measure of discordance with no need to mode filtering.

\subsection{Are WMAP and Planck that consistent one with another? A case where $N_{KL}$ does not reduce to a small number} \label{section:WMAP-Planck}

\begin{figure}[t]
\begin{center}
 {\includegraphics[width=18cm]{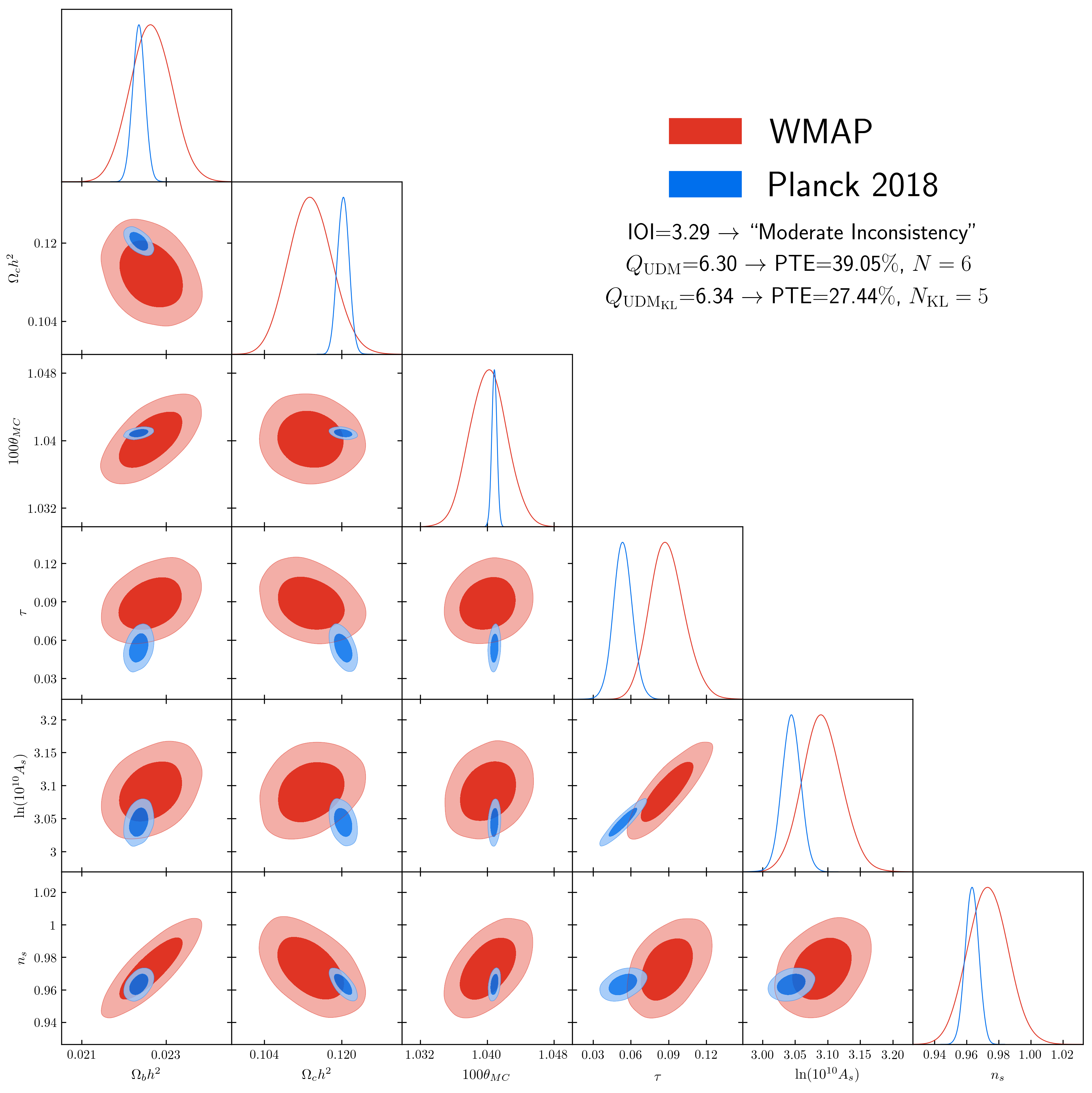}} 
\end{center}
\caption{Constraints form WMAP versus Planck-2018 data sets along with the calculated IOI and $Q_{\text{UDM}}$. For full WMAP and Planck data, the constraints are nearly Gaussian. Analyzing consistency between WMAP and Planck-2018 gives $\text{IOI}=3.29$. For $Q_{\text{UDM}}$, updating WMAP with Planck-2018 data gives $Q_{\text{UDM}}=6.30$ with a $\text{PTE}=39.05\%$. After using the KL decomposition, one gets $Q_{\text{UDM}_\text{KL}}=6.34$ and $\text{PTE}=27.44\%$. Although we provide here the 2D and 1D plots, we recall the important point that inconsistency between data sets are must considered in the full parameter spaces because marginalization can only reduce inconsistency, see e.g. \cite{WL2017a,Handley-etal-2019-suspiciousness}}
\label{Fig:WMAP-vs-Planck}
\end{figure}

\begin{figure}[t]
\begin{center}
\begin{tabular}{ c c }
 {\includegraphics[width=8cm]{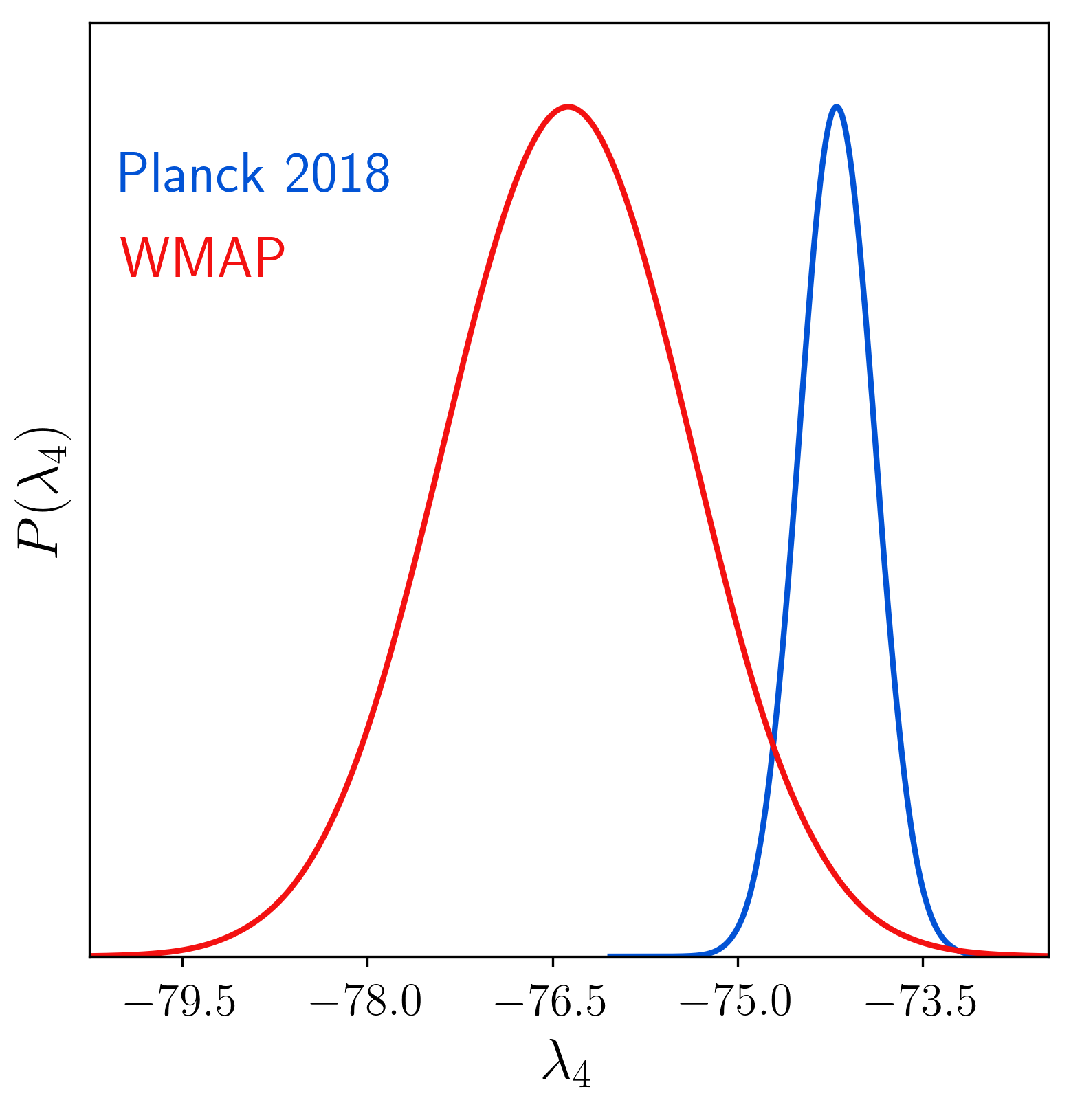}} 
 {\includegraphics[width=8cm]{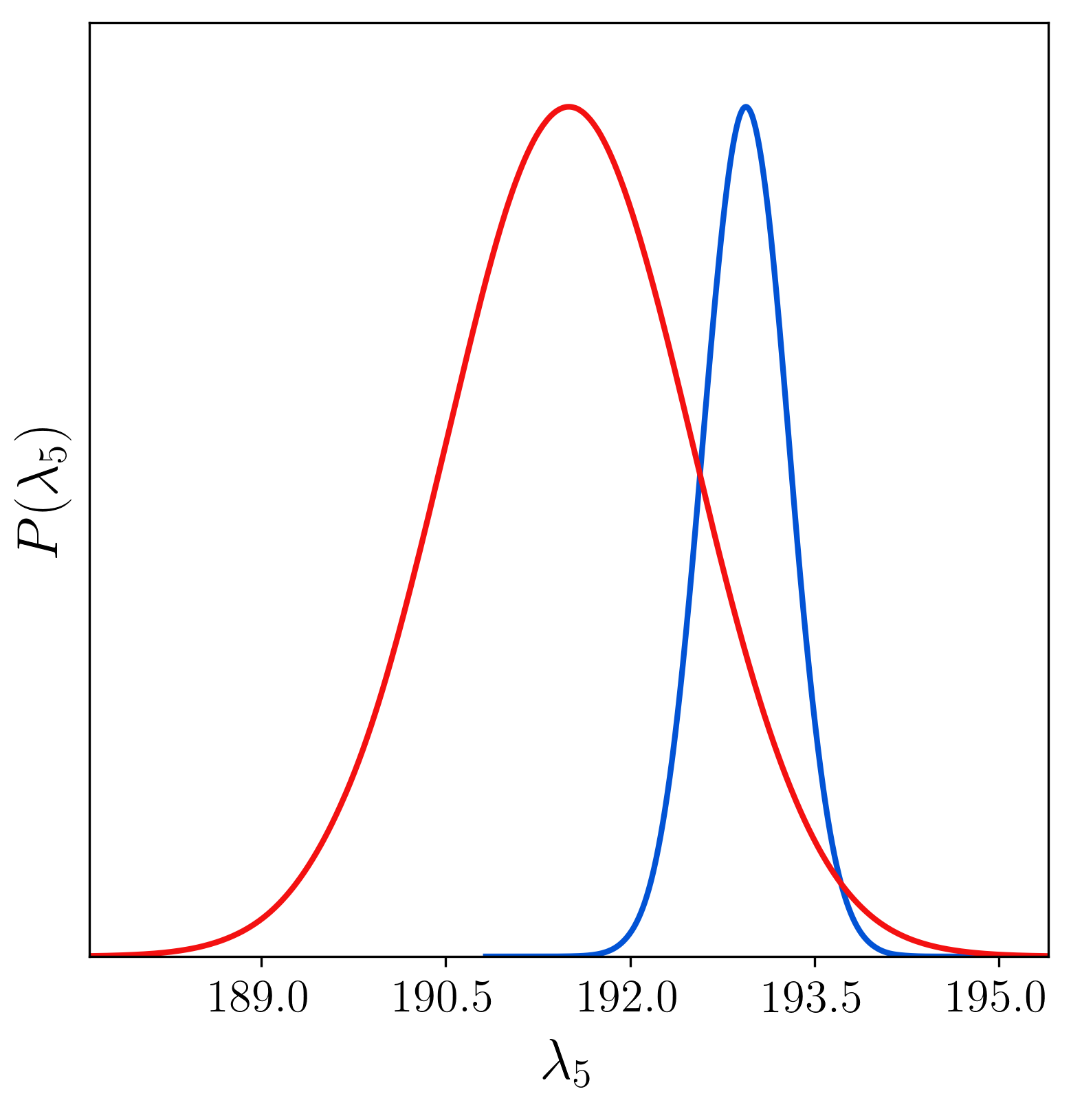}}
\end{tabular}
\end{center}
\caption{WMAP versus Planck: decorrelated modes with tension. Gaussian distributions obtained using the mean value and standard deviation for the modes with some tension after a KL decomposition using (\ref{IOI-eig0}). Shown are the fourth and fifth modes with 2.1-$\sigma$ and 1.4-$\sigma$ tension, respectively. The difference between this plot and Fig. \ref{Fig:WMAP-vs-Planck} is that here the different $\lambda$ modes are decorrelated. As we explain in the text, this helps understand the tension beyond the projected 2D contour plots.}
\label{Fig:KLmodes_WMAPvsPlanck}
\end{figure}

We recall that the number of degrees of freedom is usually reduced by the KL decomposition and the dropping of modes with numerical noise from MCMC sampling. The modes that are dominated by numerical noise come from comparing experiments where the second data set added does not improve much over the first data set (Gaussian), leading to $\lambda^\alpha \approx  1$ (see section \ref{section-QUDM}). But this does not happen in all cases of interest as we discuss here.  

Indeed, it is instructive to analyze IOI and $Q_{\text{UDM}}$ in a situation where, even after the KL decomposition, the number of degrees of freedom remains high. For that, we compare WMAP9 which provides  constraints with nearly Gaussian distribution for the base parameters against the full Planck-2018 data, which gives also Gaussian distributions for the parameters but with a higher constraining power. 

We use the nine year results full WMAP data \cite{2012-WMAP-CosmoParamsConstraints} and the full Planck-2018 data, TTTEEE high-$\ell$, the low-$\ell$ polarization used in \cite{Planck2018-CosmoParamsConstraints} and the CMB lensing measurements \cite{Planck2018-CMBlensing} in a joint analysis. Thus, we proceed by starting with WMAP data and then using the update procedure of $Q_{UDM}$ using Planck data which should offer enough improvement over each mode so no decrease in $N_{\text{KL}}$ is expected. We find that indeed, $N_{\text{KL}}=5$  when six base cosmological parameters where considered (i.e. N=6).  The analysis gives $Q_{\text{UDM}}=6.30$ and after using the KL decomposition and removing the modes with numerical noise (in this case only one mode) we obtain $Q_{\text{UDM}_{\text{KL}}}=6.34$. 

Meanwhile, the comparison between WMAP and Planck gives $\text{IOI=3.29}$. Again, we can observe that because of the nearly Gaussian distributions of the parameters in both experiments we get $\text{IOI} \approx Q_{\text{UDM}_{\text{KL}}}/2$. 

Now, interpreting the value of IOI using the Jeffrey's scale indicates a moderate inconsistency. On the other hand, interpreting $Q_{\text{UDM}}$ using PTE leads to a non-significant inconsistency with $\text{PTE}=39.05\%$ for $Q_{\text{UDM}}$ and $\text{PTE}=27.44\%$ for $Q_{\text{UDM}_{\text{KL}}}$ after removing one noise-dominated mode. However, this way of interpreting the results using PTE with a yet high number of degrees of freedom may underestimate the inconsistency as discussed in sub-section \ref{section:KLmodes} where we concur with the arguments of Ref. \cite{IOI-remarks}.  

Finally, in order to examine the tension found above by IOI, it is useful to analyze the constraints in a space where the parameters are not correlated. Therefore, we compare WMAP and Planck-2018 constraints by applying a change of basis using Eq. (\ref{IOI-eig0}) and visualize the constraints when the parameters are decorrelated. Part of the results are shown in in Fig. \ref{Fig:KLmodes_WMAPvsPlanck} where we plot the two KL-modes that present the largest tensions. Among the six different modes, four of them are below the 1-$\sigma$ level of tension. However, two modes present a tension of 1.4-$\sigma$ and 2.1-$\sigma$. This can help one to see that even when there might not be a strong tension between WMAP and Planck-2018, the data sets are not as consistent as we may think from looking at the 1D marginalized contours. It is worth mentioning that other works have also pointed out some ~3-$\sigma$ tension present between WMAP and Planck-2015 data \cite{2015ApJ...801....9L,2014PhRvD..89d3004H}.  

%%%%%%%%%%%%%%%%%%%%%%%%%%%%%%%%%%%%%%%%%%%%%%%%%%%%%%%%%%%%%%%%%%%%%%%%%%%%%%%%%%%%%%%%%%%%%%%%%%%%%%%%%%%%%%%%%%%%%%%%%%%%%%%%%%%%%%%%%%%%%%
%%%%%%%%%%%%%%%%%%%%%%%%%%%%%%%%%%%%%%%%%%%%%%%%%%%%%%%%%%%%%%%%%%%%%%%
\section{Application and results from current data sets} \label{section:application-current-data}
%%%%%%%%%%%%%%%%%%%%%%%%%%%%%%%%%%%%%%%%%%%%%%%%%%%%%%%%%%%%%%%%%%%%%%%%%%%%%%%%%%%%%%%%%%%%%%%%%%%%%%%%%%%%%%%%%%%%%%%%%%%%%%%%%%%%%%%%%%%%%%
Combining data sets from different cosmological probes is crucial to disentangle the values of the cosmological parameters in a precise way. Also important is to jointly analyze data sets that are consistent with each other, in order to obtain reliable constraints. Therefore, after verifying the numerical stability of IOI, we apply it in order to trace inconsistencies between current data sets in this section. We compare CMB measurements from Planck-2018 against different Large-Scale-Structure (LSS) measurements in combination with some added probes. Additionally, we test the consistency between different measurements of the Hubble constant using IOI.
%%%%%%%%%%%%%%%%%%%%%%%%%%%%%%%%%%%%%%%%%%%%%%%%%%%%%%%%%%%%%%%%%%%%%

%%%%%%%%%%%%%%%%%%%%%%%%%%%%%%%%%%%%%%%%%%%%%%%%%%%%%%%
\subsection{Planck-2018 versus Large-Scale-Structure data sets}
\begin{table}[b]
\caption{List of data sets used in order to compare Planck-2018 against LSS and different $H_0$ measurements. We also list some additional probes added in order to break some degeneracies between some cosmological parameters and obtain Gaussian parameter distributions when combined with LSS data sets.}
\renewcommand{\arraystretch}{1.2}
\begin{tabular}{ m{3cm}  m{10cm}  l }
\hline\hline
Data sets & \centering{Description} & Added probes \\ [3pt] \hline

\multicolumn{3}{l}{1. Background}  \\ \hline
SH0ES & Locally measured Hubble constant \cite{2019-Riess-H0} & N/A \\
H0LiCOW & Strong gravitational lensing \cite{2019-H0LiCOW-XIII} & N/A\\
CCHP-TGRB\footnote{CCHP stands for Carnegie-Chicago Hubble Program} &  Tip of the Red Giant Branch (TRGB) applied to SNe Ia \cite{CCHP-2019-Freedman} & N/A\\
TRGB-2 & TRGB+SNe Ia distance ladder as re-analyzed with different methods by \cite{Yuan-etal-2019-TRGB-local} & N/A \\
SNe  & SNe Ia joint analysis data from Pantheon compilation \cite{Pan-STARRS-2017}  & BBN\footnote{Primordial deuterium abundance $D/H=2.527 \pm 0.030 \times 10^{-5}$ \cite{2017Cooke}.} \\ 
BAO & \multirow{3}{8.8cm}{Six Degree Field Galactic Survey (6dF) ($z_{eff}=0.106$) \cite{2011BAO-6df}, SDSS main galaxy sample (MGS) ($z_{eff}=0.15$) \cite{2015BAO-sdss-mgs} and BAO consensus constraints \cite{AlamEtAl2016}} & BBN \\ 
 & & \\
 & & \\\hline

\multicolumn{3}{l}{2. Planck-2018 temperature and polarization} \\ \hline
TT+lowE  & Planck high-$\ell$\footnote{High-$\ell$ represents the range $30\leq\ell\leq 2058$.} temperature auto correlation \cite{2018-Planck-cosmo-params}  & lowE\footnote{Planck-2018 low-$\ell$ ( $2\leq\ell\leq 29$) polarization \cite{2018-Planck-cosmo-params}. Opposite to our previous work, here we only use the $EE$ likelihood since a poor statistical consistency of the $TE$ spectrum was reported in \cite{2018-Planck-cosmo-params}.} \\
TE+lowE  & Planck high-$\ell$ temperature-E polarization cross correlation \cite{2018-Planck-cosmo-params}  & lowE \\
EE+lowE  & Planck high-$\ell$ E-mode polarization auto correlation \cite{2018-Planck-cosmo-params}  & lowE \\
TTTEEE+lowE  & Planck high-$\ell$ temperature and E-mode polarization joint data set \cite{2018-Planck-cosmo-params}  & lowE \\
\hline

\multicolumn{3}{l}{3. Large-Scale-Structure data sets} \\ \hline
DES  & Dark Energy Survey Year 1 clustering and lensing analysis \cite{2017-DESC-1st-joint} & lowE+SNe+BBN \\
CMB lens  & Planck-2018 CMB lensing measurements \cite{Planck18-CMBlens}  & lowE+SNe+BBN \\
SDSS RSD  & SDSS III galaxy clustering data from BAO spectroscopic survey \cite{AlamEtAl2016} & lowE+SNe+BBN \\
Joint LSS  & Joint analysis combining DES+CMB lens+SDSS RSD & lowE(or with priors)+BBN \\
\hline\hline
\end{tabular}
\label{Table:DataSets}
\end{table}

Inconsistencies between Planck and LSS have been pointed out in previous works \cite{2019arXiv190105289D,2019EPJC...79..576K,2019-tensions-WillHandley,2013CFHTlens,2017-KiDS-Weak-lensing,2011WiggleZ-growth-rate,WL2017b,2014sdssIII-redshift-space,2015Planck-SZ-cluster-count,Planck2018-CosmoParamsConstraints}. Here, we do a discordance/concordance analysis between the recently released Planck-2018 data against different LSS data sets using IOI. 

We list the data sets used in this analysis in Table \ref{Table:DataSets}. In order to break denegeracies between some cosmological parameters and obtain Gaussian distribution for the parameters, we use some data sets as added probes in combination with the LSS measurements. However, we do this while taking care of not affect any inconsistency, if present, by such additions. 

We consider three different data sets for LSS measurements. As we can see in Table \ref{Table:LSS_vs_LSS}, all data sets are consistent one with each other since IOI is less than the unity in each case. Moreover, the multi-experiment IOI is also less than unity when all LSS measurements are combined. Then, this allows us to combine them in a joint analysis as shown in section \ref{section:IOI-H0}. However, it is important to mention that in order to get nearly Gaussian distributions for all the parameters, we use lowE+SNe+BBN as added probes. Similarly, we can repeat the same analysis for the internal consistency between  Planck temperature and polarization measurements as shown in Table \ref{Table:Planck_vs_Planck}. In this case, we find that the values of IOI for TT+lowE when compared to TE+lowE or EE+lowE show an inconsistency in the high end of the weak range and close to 2.5. Furthermore, the values of IOI shows that a moderate inconsistency is present for TE+lowE and EE+lowE. Similarly, we find that the multi-experiment IOI shows a moderate inconsistency when combining the three data sets. Therefore, some moderate internal inconsistency is found in the Planck-2018 data indicating perhaps the presence of some systematic effects. 

Next, we compare the Planck data against the LSS measurements after analyzing the self-consistency of Planck and LSS.The Planck temperature and polarization data shows a weak inconsistency with CMB lensing and SDSS RSD measurements. Therefore, we consider that combining these experiments in a joint analysis is fine, but we point out that further improvements need to be done in order to mitigate these small inconsistencies, even when they are weak inconsistencies. On the other hand, DES data shows a moderate inconsistency with TT+lowE, EE+lowE and TTTEEE+lowE data sets. This moderate inconsistency with DES seems similar to what was reported in \cite{Planck2018-CosmoParamsConstraints} when including galaxy clustering data in DES data, as moderate percent-level tension. Additionally, we show that there is a moderate inconsistency between joint LSS and TE+lowE, as well as for EE+lowE. However, this inconsistency becomes a strong inconsistency when comparing joint LSS with TT+lowE or TTTEEE+lowE. Therefore, even when there is a weak inconsistency between Planck and CMB lens or SDSS RSD, or a moderate inconsistency with DES, combining the LSS probes in a joint analysis seriously increases the tension with Planck. It is important to mention that, the strong inconsistency between Planck and joint LSS should be attributed to LSS measurements and not BBN, since deuterium abundance determination \cite{2017Cooke} was found to be in no significant tension with Planck-2018 data \cite{Planck2018-CosmoParamsConstraints}. Then, we point out that one should be aware of this tension when combining Planck and different LSS probes in a joint analysis. 

It remains to be explored whether these inconsistencies are due to systematic effects in any of the data sets. However, we find that while SDSS RSD and CMB lensing show smaller inconsistencies with Planck-2018 (temperature and polarization data) which are lower with respect to what was found when comparing SDSS RSD and CMB lensing versus Planck-2015 in previous study \cite{WL2017b}.  However, the inconsistency between weak lensing and Planck persist to be a moderate inconsistency, according to Table \ref{Table:JeffreyScale}. Finally, the inconsistency with a joint analysis of this LSS probes and Planck is a strong inconsistency with IOI=5.27.

\begin{table}[t]
\renewcommand{\arraystretch}{1.2}
\begin{tabular}{ p{2cm} | >{\centering}m{2cm} >{\centering}m{2cm} r }
\hline\hline
    & CMB lens & DES & SDSS RSD \\ \hline
CMB lens & - & 0.41 & 0.75 \\
DES & 0.41 & - & 0.81 \\
SDSS RSD & 0.75 & 0.81 & - \\ \hline
\multicolumn{3}{l}{Multiexperiment IOI = 0.74} \\
\hline\hline
\end{tabular}
\caption{Two-experiment IOI values for comparisons between LSS data sets. As mentioned in Table \ref{Table:DataSets}, we add the combination lowE+SNe+BBN to each LSS data set in order to break degeneracies between parameters. As shown, all values of IOI are below unity, even for the multi-experiment IOI. Then, according to Table \ref{Table:JeffreyScale}, {we find that all LSS data sets are consistent with one another}.}
\label{Table:LSS_vs_LSS}
\end{table}

\begin{table}[t]
\renewcommand{\arraystretch}{1.2}
\begin{tabular}{ p{2cm} | >{\centering}m{2cm} >{\centering}m{2cm} r }
\hline\hline
  Planck  & TT+lowE & TE+lowE & EE+lowE \\ \hline
TT+lowE & - & 2.43 & 2.45 \\
TE+lowE & 2.43 & - & 3.19 \\
EE+lowE & 2.45 & 3.19 & - \\ \hline
\multicolumn{3}{l}{Multiexperiment IOI = 3.39} \\
\hline\hline
\end{tabular}
\caption{Two-experiment IOI values for self-consistency analysis between Planck temperature and polarization data sets. TT+lowE shows a weak inconsistency between TE+lowE and EE+lowE. {EE+lowE and TE+lowE show a moderate inconsistency}.}
\label{Table:Planck_vs_Planck}
\end{table}

\begin{table}[t!]
\renewcommand{\arraystretch}{1.2}
\begin{tabular}{ p{3cm} | >{\centering}m{2cm} >{\centering}m{2cm} >{\centering}m{2cm} >{\raggedleft\arraybackslash}m{1.7cm} }
\hline\hline
    & CMB lens & DES & SDSS RSD & Joint LSS \\ \hline
TT+lowE & 1.83 & 3.28 & 0.90 & 5.01 \\
TE+lowE & 1.54 & 1.54 & 0.40 & 3.90 \\
EE+lowE & 2.07 & 3.13 & 1.39 & 3.97 \\
TTTEEE+lowE & 1.60 & 3.14 & 0.70 & 5.27 \\
\hline\hline
\end{tabular}
\caption{Two-experiment IOI values for Planck temperature and polarization data against LSS data sets. As shown, there is is a weak inconsistency between Planck data and CMB lensing data. Similarly when comparing Planck against SDSS RSD data. However, DES shows a moderate inconsistency with Planck TT+lowE, EE+lowE and TTTEEE+lowE. The largest tension is between DES and TT+lowE with IOI$\approx$3.5 which is similar to a previous study comparing CFHTLens and Planck-2015 \cite{WL2017b}. In order to calculated these results, we used our own chains for all results but we double-checked them using the chains from Planck-2018 \cite{Planck2018-CosmoParamsConstraints}. The results were found in agreement as listed above. {We find strong tension between Planck and joint LSS probes with IOI=5.27}.}
\label{Table:Planck_vs_LSS}
\end{table}

%%%%%%%%%%%%%%%%%%%%%%%%%%%%%%%%%%%%%%%%%%%%%%%%%%%%%%%

%%%%%%%%%%%%%%%%%%%%%%%%%%%%%%%%%%%%%%%%%%%%%%%%%%%%%%%
\subsection{Consistency analysis of the Hubble constant from various data sets} \label{section:IOI-H0}

\begin{figure}[t]
\begin{center}
 {\includegraphics[width=18cm]{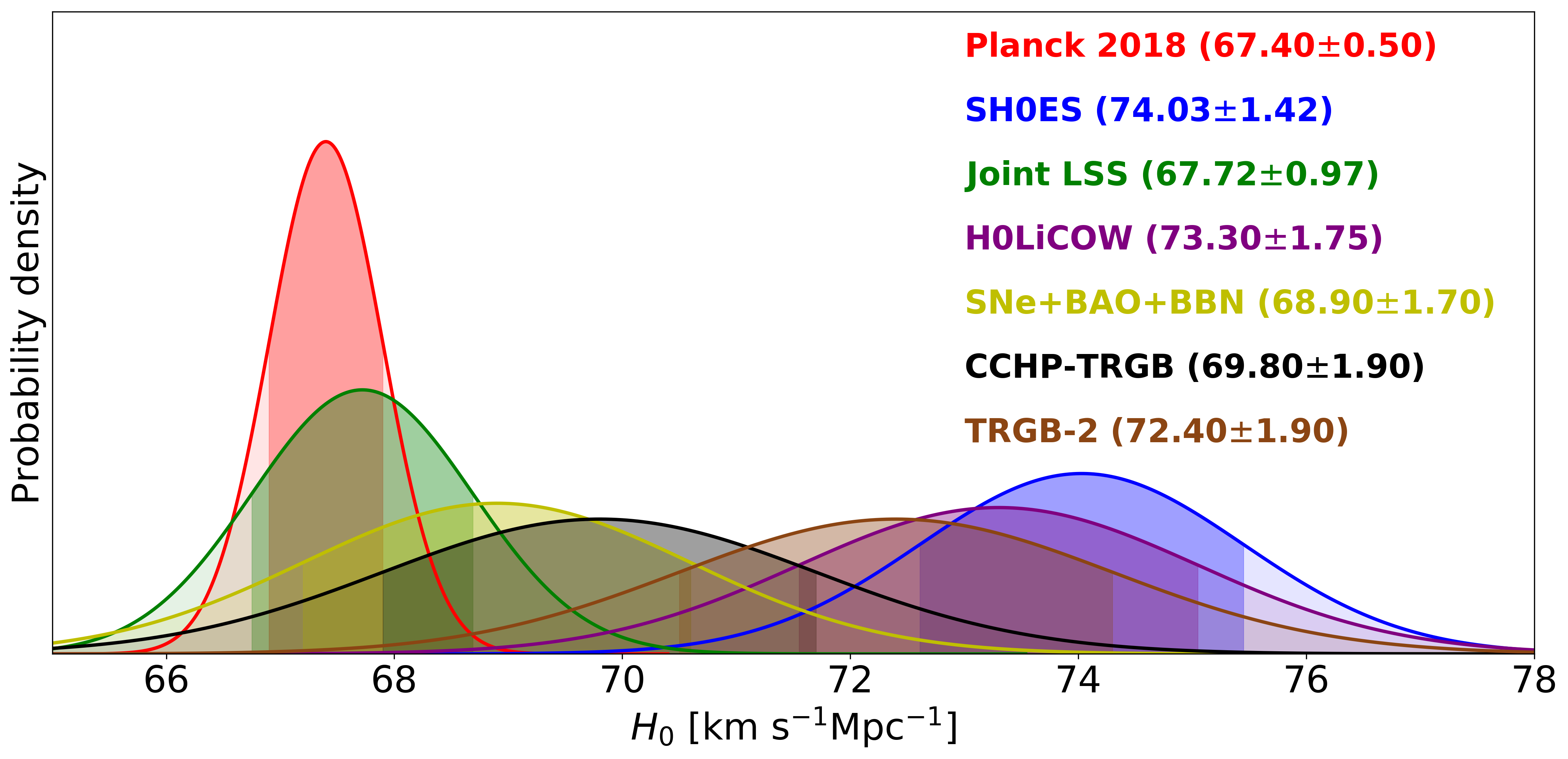}} 
\end{center}
\caption{One dimensional probability distributions for $H_0$ given the current measurements. The shaded area represents the 1-$\sigma$ confidence region of each probability distribution function.}
\label{Fig:H0_measurements}
\end{figure}

%\begin{table}[t]
%\renewcommand{\arraystretch}{1.2}
%\begin{tabular}{ m{2.2cm} | >{\centering}m{2.2cm} >{\centering}m{2.2cm} >{\centering}m{2.2cm} >{\centering}m{2.2cm} >{\centering}m{2.2cm} >{\centering}m{2.2cm} >{\raggedleft\arraybackslash}m{1.cm} }
%\hline\hline
%Methods &  Planck & SH0ES & Joint LSS (with priors) &  H0LiCOW  &  SNe+BAO+BBN & CCHP & TRGB-2 \\ \hline
%$H_0 \left(\frac{\text{km/sec}}{\text{Mpc}}\right)$ & $67.4\pm0.50$ & $74.03\pm1.42$  & $67.72 \pm 0.97$ & $73.30\pm1.75$ & $68.90\pm1.70$ & $69.80\pm1.90$ & $72.4\pm1.90$  \\
%\hline\hline
%\end{tabular}
%\caption{Constraints on $H_0$ from six different experiments.}
%\label{Table:H0-measurements}
%\end{table}

\begin{table}[t]
\renewcommand{\arraystretch}{1.2}
\begin{tabular}{ m{2.6cm} | >{\centering}m{2.1cm} >{\centering}m{2.1cm} >{\centering}m{2.1cm} >{\centering}m{2.1cm} >{\centering}m{2.1cm} >{\centering}m{2.1cm} >{\raggedleft\arraybackslash}m{1.5cm} }
\hline\hline
 &  Planck & SH0ES & Joint LSS (with priors) &  H0LiCOW  &  SNe+BAO +BBN & CCHP-TRGB & TRGB-2 \\ \hline
Planck & $-$  & 9.70  & 0.04 & 5.25 & 0.36 & 0.75 & 3.24 \\
SH0ES\footnote{We consider SH0ES, CCHP-TRGB and TRGB-2 as representing the same probe.} & 9.70  & $-$ & 6.73 & 0.05  & 2.68 & 1.59 & 0.24 \\
Joint LSS & 0.04 & 6.73 & $-$ & 3.89 & 0.18 & 0.48 & 2.41 \\
H0LiCOW & 5.25  & 0.05  & 3.89 &  $-$ & 1.63 & 0.92 & 0.06 \\
SNe+BAO+BBN & 0.36 & 2.68  & 0.18 & 1.63 & $-$ & 0.06 & 0.94 \\
CCHP-TRGB$^{a}$  & 0.75 & 1.59  & 0.48 & 0.92 & 0.06 & $-$ & 0.47 \\
TRGB-2$^{a}$ & 3.24 & 0.24 & 2.41 & 0.06 & 0.94 & 0.47 & $-$ \\
\hline\hline
\end{tabular}
\caption{Two-experiment IOI values for the seven data sets used to measure $H_0$. While joint LSS, SNe+BAO+BBN and Planck seem consistent with each other, H0LiCOW and SH0ES measurements show a strong inconsistency with Planck. In particular, Planck against SH0ES measurements give a strong IOI=9.70 can be converted in 1D to $4.4$-$\sigma$. This is similar to what was reported in \cite{2019-Riess-H0}. Finally, since H0LiCOW and the local measurement of $H_0$ are consistent one with another. In this case, H0LiCOW+SH0ES versus Planck gives IOI=14.03 or equivalently in 1D a 5.3-$\sigma$ tension, in agreement with \cite{2019-H0LiCOW-XIII}.
}
\label{Table:H0-IOI-results}
\end{table}

%\begin{table}[t!]
%\renewcommand{\arraystretch}{1.2}
%\begin{tabular}{ m{3cm} | >{\centering}m{1.2cm} >{\centering}m{1.8cm} >{\centering}m{1.8cm} >{\centering}m{1.8cm} >{\centering}m{1.8cm} >{\centering}m{1.8cm} >{\centering}m{1.8cm}  >{\raggedleft\arraybackslash}m{1.6cm} }
%\hline\hline
%Data sets & All & Removing Planck & Removing SH0ES & Removing Joint LSS (with priors) & Removing H0LiCOW & Removing SNe+BAO +BBN & Removing CCHP-TRGB & Removing TRGB-2 \\ \hline
%Mutiexperiment IOI & 4.69 & 3.15 & 2.84 & 5.31 & 4.21 & 5.47 & 5.40 & 4.78 \\
%\hline\hline
%\end{tabular}
%\caption{Multi-experiment IOI values from seven different measurements of $H_0$. Furthermore, we compute the multi-experiment IOI when removing each of the data sets from the whole set of seven experiments.}
%\label{Table:H0-IOI-multiexp}
%\end{table}

\begin{table}[t!]
\renewcommand{\arraystretch}{1.2}
\begin{tabular}{ m{4cm} | >{\centering}m{1.5cm} >{\centering}m{2.2cm} >{\centering}m{2.2cm} >{\centering}m{2.2cm} >{\centering}m{2.2cm} >{\raggedleft\arraybackslash}m{2.5cm} }
\hline\hline
Data sets & All & Removing Planck & Removing SH0ES & Removing joint LSS (with priors) & Removing H0LiCOW & Removing SNe+BAO+BBN  \\ \hline
Mutiexperiment IOI & 5.63 & 4.38 & 2.73 & 6.90 & 4.94 & 7.01 \\
\hline\hline
\end{tabular}
\caption{Multi-experiment IOI values from five different measurements of $H_0$. Furthermore, we compute the multi-experiment IOI when removing each of the data sets from the whole set of five experiments.}
\label{Table:H0-IOI-multiexp}
\end{table}

The clear tension reported in the Hubble constant between the local measurement and the CMB inferred value has generated a lot of discussions and comparison, see for example the partial list \cite{2019-Riess-H0,2019-Vagnozzi-H0tension, 2019arXiv190512618L, 2019arXiv190703778D, 2018PhRvD..97l3504P, 2019PhRvD..99d3519G, 2019ApJ...871..210E, 2019PhRvD.100b3505C, 2019arXiv190200534K, 2017JHEP...09..033P, 2017-Zhao-etal-nature-astronomy, 2017EPJC...77..418F, 2017PhRvD..96h3513D, 2017PhRvD..96d3503D, 2017PhLB..774..317S, 2019PhRvD..99j3526K, 2018-Bolejko.K, 2016-Bernal-Verde-Riess-H0, 2016-Valentino-Melchiorri-Silk-reconciling-Planck-local}. 
In our previous work, we presented five different methods to constraint $H_0$ \cite{WL2017b}. Here, we update the analysis with current new data sets and the way we combine them. 

We use the data sets presented in Table \ref{Table:DataSets} to compare different $H_0$ measurements from \textbf{1) Planck:} CMB measurements using Planck-2018 data, specifically we use the measurement TTTEEE+lowE+CMB lens used in \cite{Planck2018-CosmoParamsConstraints}; \textbf{2) SH0ES:} Local measurements of $H_0$ from \cite{2019-Riess-H0}; \textbf{3) Joint LSS:} Joint analysis using three distinct LSS measurements as in Table \ref{Table:DataSets} \footnote{In addition, we add BBN in order to break the degeneracy in $\Omega_b h^2$.}; \textbf{4) SNe+BAO+BBN:} Combination of background data sets to constraint $H_0$. SNe and BAO have different degeneracy directions, while BBN can constraint $\Omega_b h^2$; \textbf{5) H0LiCOW:} measures expansion rate of the universe using time-delay cosmography and distance ladder results \cite{2019-H0LiCOW-XIII}; \textbf{6) CCHP-TRGB:} Calibration of SNe Ia using the TRGB method, independent of the Cepheid distance scale \cite{CCHP-2019-Freedman}; \textbf{7) TRGB-2:} TRGB+SNe Ia distance ladder using a different calibration method  \cite{Yuan-etal-2019-TRGB-local}. The values of $H_0$ obtained from each of these methods are given in Fig. \ref{Fig:H0_measurements}.

We report the values of IOI for all the combinations of the seven data sets in Table \ref{Table:H0-IOI-results}. We can see that the biggest inconsistency is obtained when comparing Planck against SH0ES measurements, leading to a value of IOI=9.70 which is a strong inconsistency. In one dimension (one parameter), we can convert this IOI value to the commonly used confidence level of tension by using $n$-$\sigma=\sqrt{2\text{IOI}}$. Then, we obtain a tension between Planck-2018 and SH0ES measurements of $4.4$-$\sigma$. This is exactly what was reported in \cite{2019-Riess-H0}. Similarly, Planck and H0LiCOW indicate a strong inconsistency with IOI higher than 5. Here, it is worth mentioning that SH0ES and H0LiCOW measurements are consistent one with each other with a value of IOI=0.05. Therefore, both data sets can be jointly analyzed. A previous work reported that by combining SH0ES and H0LiCOW data sets in a joint analysis gives $H_0=73.8\pm1.1 $km s$^{-1}$Mpc$^{-1}$ \cite{2019-H0LiCOW-XIII}, producing a tension of 5.3-$\sigma$ with Planck. In our case, we obtain IOI=14.03 if we compare H0LiCOW+SH0ES against Planck. This is equivalent to the 5.3-$\sigma$ level of tension mentioned before. Similarly, a strong inconsistency is present between joint LSS and late universe probes H0LiCOW+SH0ES with IOI=8.59, equivalent to a 4.1-$\sigma$ level of tension.

 We find that joint LSS presents a strong inconsistency with SH0ES and a moderate inconsistency with H0LiCOW, which can be translated to a tension of 3.7-$\sigma$ and 2.8-$\sigma$, respectively. Furthermore, SNe+BAO+BBN presents a moderate inconsistency with SH0ES but a weak inconsistency with H0LiCOW. On the other hand, Planck seems to be consistent with joint LSS and SNe+BAO+BBN, while these two data sets are actually consistent one with another. 
 
 The recent Hubble constant measurement obtained by \cite{CCHP-2019-Freedman} using the Tip of the Red Giant Branch is midway and in overall consistent with both SH0ES and Planck, with only an IOI of 1.59 with SH0ES (1.78-$\sigma$) and an IOI of 0.75  1.23-$\sigma$ with Planck, again in agreement with the values reported in \cite{CCHP-2019-Freedman}.
 Ref. \cite{Yuan-etal-2019-TRGB-local} redid the TRGB analysis using other correction/calibration methods and found a higher value of $H_0$, still in some tension with Planck. 
 
We also compute the multi-experiment IOI values using Eq. (\ref{IOIeq:mult-exp}) in Table \ref{Table:H0-IOI-results}. Comparing all data sets leads to values of IOI$=5.63$, which stands for a strong inconsistency according to Table \ref{Table:JeffreyScale}. We remove alternatively some data sets and see how the inconsistency drops. This value goes higher if we remove SNe+BAO+BBN or joint LSS , making the inconsistency more severe. This is because 1) these data sets are consistent one with another, and 2) these data sets are consistent with Planck. On the other hand, the value of the multi-experiment IOI drops if we remove Planck, H0LiCOW or SH0ES measurements. Indeed, removing these data sets reduce the inconsistency to a moderate inconsistency.

Noteworthy is that we find here a new strong tension between SH0ES and that of joint LSS probes with IOI=6.73 corresponding in 1D to 3.7-$\sigma$. 

%%%%%%%%%%%%%%%%%%%%%%%%%%%%%%%%%%%%%%%%%%%%%%%%%%%%%%%

%%%%%%%%%%%%%%%%%%%%%%%%%%%%%%%%%%%%%%%%%
\section{Summary and concluding remarks} \label{section-conclusions}
%%%%%%%%%%%%%%%%%%%%%%%%%%%%%%%%%%%%%%%%%

First, we studied some properties of recently proposed parameter-distance based measures of inconsistencies. Namely the index of inconsistency (IOI) of Ref. \cite{WL2017a} and the updated measure ($Q_{\text{UDM}}$) of Ref. \cite{2018-Raveri-Hu}. We showed that in the nearly Gaussian cosmological parameter distributions and weak prior limit, the two measures are equal, up to a factor 2, i.e. $Q_{\text{UDM}}=2\,\text{IOI}$. A Karhunen-Loeve (KL)  decomposition was applied to $Q_{\text{UDM}}$ in \cite{2018-Raveri-Hu} in order to eliminate modes with numerical noise for the $Q_{\text{UDM}}$ quantity. We showed that IOI does not need such a decomposition. 

Importantly, we investigate the question of stability by repeating cosmological and consistency analyses using the two measures, imposing two convergence limits, and reporting the mean values and standard deviations of the results obtained. We find that IOI gives a small relative standard deviation of $2.14\%$ which indicates high stability. $Q_{\text{UDM}}$ gives a relative standard deviation of $11.08\%$ without the KL decomposition and $4.14\%$ after such a decomposition. It is thus found here that IOI gives a better standard deviation than $Q_{\text{UDM}}$ with or without KL decomposition. Our finding was also confirmed by an analytical result using KL decomposition.

Next, we discussed the question of interpretation of these inconsistency measures and the effect of the KL decomposition and its mode-filtering algorithms. It was argued in a recent analysis \cite{IOI-remarks} that the procedure of translating the values of IOI or $Q_{\text{UDM}}$ into probability to exceed and significance level in high dimensional (parameter) spaces can underestimate inconsistencies due to high number of parameters or degrees of freedom. It was explain there that while such a procedure is justified when performing data fitting, importing it to the interpretation levels of inconsistency measures may not be the case. 

Specifically, we showed here that while the KL decomposition in general transforms the degree of freedom from $N$ to $N_{KL}$ which is much smaller than $N$, this is not always the case. We provided an illustrative construction where $N$ stays the same and also showed that in the case of WMAP updated by Planck it decreases from $N=6$ to $N_{KL}=5$. So the effect of underestimating inconsistency may still happen here.

An example was also provided where the mode-filtering algorithm used with the KL decomposition discards a mode that contains a significant inconsistency and thus misses it.

In a second part of the paper, we perform various consistency analyzes using IOI between current multiple data sets and \textit{working within the entire parameter spaces}. 

We find current LSS data sets (CMB lensing from Planck-2018, DES Y1 and SDSS RSD DR12) to be all consistent one with another with IOI less than 1. 
This is not the case for Planck temperature (TT) vs polarization (TE,EE) data where moderate inconsistencies are present, particularly  between  TE+lowE and EE+lowE with IOI=3.19 and when all combined with IOI=3.39. 
Moderate inconsistencies are also found for Planck-2018 versus DES-Y1 with IOI=3.28. Such an inconsistency reaches the strong range when comparing Planck versus joint LSS probes with IOI=5.01 and IOI=5.27 when using TT+lowE and TTTEEE+lowE, respectively. Our results thus indicate that moderate and non-ignorable tensions are present between Planck and LSS when the whole parameter spaces are compared. This is in agreement with previous studies that focused on marginalized constraints on $S_8$ or $\Sigma_8$ such as in  Refs. \cite{2019-Wibking,2015-Dossett-etal-Planck-CFHTlens,2019-Joudaki-Kids+DES}.  Unlike previous results in \cite{WL2017b,2018-Raveri-Hu} using Planck-2015, we find here that CMB lensing has only weak inconsistencies with CMB temperature and polarization data from Planck-2018.  

Finally, we provided a consistency study of the Hubble constant as determined from seven data sets. We find and confirm a strong inconsistency between the local measurement from SH0ES (LMH) versus Planck-2018 with IOI=9.70 (i.e. 4.40-$\sigma$ in this 1D case) and concurring with \cite{2019-Riess-H0}. Next, H0LiCOW versus Planck gives IOI=5.25 (3.24-$\sigma$). Moreover, since LMH and  H0LiCOW  are consistent one with another (IOI=0.05), we can combine them and compare them to Planck. We find that H0LiCOW+LMH versus Planck gives IOI=14.03 (i.e. 5.30-$\sigma$) which is substantial and in agreement with \cite{2019-H0LiCOW-XIII}. We also compared the constraints from CCHP-TRGB with other data sets and find them to be in overall consistency with SH0ES and Planck as reported in \cite{Freedman-etal-2019}. 

We find that joint LSS data sets (CMB lensing + DES + SDSS RSD) versus SH0ES LMH shows a new strong tension of IOI=6.73 (i.e. 3.7-$\sigma$). Similarly, a strong inconsistency is found between joint LSS and late universe probes H0LiCOW+SH0ES with IOI=8.59, equivalent to a 4.1-$\sigma$ level of tension.

The tensions discussed above are not fading away. Whether they are due to systematic effects in the data sets or the manifestation of problems with the underlying model, the causes of these old and new tensions need to be identified and addressed.

\begin{acknowledgements}
The authors thank Adam Riess for useful comments on the manuscript. MI acknowledges that this material is based upon work supported in part by the U.S. Department of Energy, Office of Science, under Award Number DE-SC0019206 and the National Science Foundation under grant AST-1517768. CGQ gratefully acknowledges a PhD scholarship from the Mexican National Council for Science and Technology (CONACYT). 

The authors acknowledge the Texas Advanced Computing Center (TACC) at The University of Texas at Austin for providing HPC resources that have contributed to the research results reported within this paper. URL: http://www.tacc.utexas.edu.

\end{acknowledgements}

\appendix

\section{Proof for $2\text{IOI}=Q_{\text{UDM}}$ in the Gaussian and weak prior limit} \label{Appendix:2IOI=QUDM}

We have shown different situations where $Q_{\text{UDM}}=2\text{IOI}$ or, using real data, that $Q_{\text{UDM}} \approx 2\text{IOI}$. However, as mentioned, this relation holds in the Gaussian and weak prior limit. Then, we proof that $Q_{\text{UDM}}$ can be derived from IOI in such a limit. Indeed, in general the joint mean value given data sets 1 and 2, is given by

\begin{equation}
    \bm{\mu_{12}} = (\bm{C_1}^{-1}+\bm{C_2^}{-1}+\bm{P}^{-1})^{-1}(\bm{C_{1}}^{-1}\bm{\mu_{1}} + \bm{C_{2}}^{-1}\bm{\mu_{2}} + \bm{P}^{-1}\bm{\mu_{P}})
    \label{AppendixEq:joint-mean-withprior}
\end{equation}
and the joint covariance matrix can be obtained by
\begin{equation}
    \bm{C_{12}} = (\bm{C_{1}}^{-1}+\bm{C_{2}}^{-1} + \bm{P}^{-1})^{-1},
    \label{AppendixEq:joint-cov-withprior}
\end{equation}
where $\bm{P}$ is the prior covariance matrix and $\bm{\mu_{P}}$ is the prior mean value. 

Now, since IOI is designed to work with Gaussian data sets we assume we are working with Gaussian distributions. Moreover, we use a weak prior limit approximation by considering $\bm{P} \rightarrow 0$. Under these assumptions, (\ref{AppendixEq:joint-mean-withprior}) and (\ref{AppendixEq:joint-cov-withprior}) can be written as
\begin{equation}
    \bm{\mu_{12}} = (\bm{C_1}^{-1}+\bm{C_2^}{-1})^{-1}(\bm{C_{1}}^{-1}\bm{\mu_{1}} + \bm{C_{2}}^{-1}\bm{\mu_{2}})
    \label{AppendixEq:joint-mean}
\end{equation}
and
\begin{equation}
    \bm{C_{12}} = (\bm{C_{1}}^{-1}+\bm{C_{2}}^{-1})^{-1},
    \label{AppendixEq:joint-cov}
\end{equation}
respectively. Then, by using matrix properties we can show that
\begin{equation}
\begin{split}
(\bm{C_{1}}^{-1}+\bm{C_{2}}^{-1})^{-1} & = \bm{C_{1}}(\bm{C_{1}}+\bm{C_{2}})^{-1}\bm{C_{2}}, \\
& = \bm{C_{2}}(\bm{C_{1}}+\bm{C_{2}})^{-1}\bm{C_{1}}.
\end{split}
\end{equation}
Then, we can write the difference of the mean value of the first data set and the mean value of the joint analysis as
\begin{equation}
\begin{split}
\bm{\mu_{1}}-\bm{\mu_{12}} & =\bm{\mu_{1}}-(\bm{C_{1}}^{-1}+\bm{C_{2}}^{-1})^{-1}(\bm{C_{1}}^{-1}\bm{\mu_{1}}+\bm{C_{2}}^{-1}\bm{\mu_{2}}), \\
& = \bm{\mu_{1}}-[\bm{C_{2}}(\bm{C_{1}}+\bm{C_{2}})^{-1}\bm{\mu_{1}}+\bm{C_{1}}(\bm{C_{1}}+\bm{C_{2}})^{-1}\bm{\mu_{2}}], \\
& =(\bm{C_{1}}+\bm{C_{2}})(\bm{C_{1}}+\bm{C_{2}})^{-1}\bm{\mu_{1}}-[\bm{C_{2}}(\bm{C_{1}}+\bm{C_{2}})^{-1}\bm{\mu_{1}}+\bm{C_{1}}(\bm{C_{1}}+\bm{C_{2}})^{-1}\bm{\mu_{2}}], \\
& = \bm{C_{1}}(\bm{C_{1}}+\bm{C_{2}})^{-1}(\bm{\mu_{1}}-\bm{\mu_{2}}).
\end{split}
\label{AppendixEq:means-simplified}
\end{equation}
Similarly, we can rewrite the difference between the covariance matrix of the first data set and the joint covariance matrix as

\begin{equation}
\begin{split}
 (\bm{C_{1}}-\bm{C_{12}})^{-1} & =[\bm{C_{1}}-(\bm{C_{1}}^{-1}+\bm{C_{2}}^{-1})^{-1}]^{-1}, \\
 &  =[(\bm{C_{1}}+\bm{C_{2}})(\bm{C_{1}}+\bm{C_{2}})^{-1}\bm{C_{1}}-\bm{C_{2}}(\bm{C_{1}}+\bm{C_{2}})^{-1}\bm{C_{1}}]^{-1}, \\
& =[\bm{C_{1}}(\bm{C_{1}}+\bm{C_{2}})^{-1}\bm{C_{1}}]^{-1}, \\
& = \bm{C_{1}}^{-1}(\bm{C_{1}}+\bm{C_{2}})\bm{C_{1}}^{-1}.
\end{split}
\label{AppendixEq:cov-simplified}
\end{equation}

Finally, we can substitute the expressions obtained in (\ref{AppendixEq:means-simplified}) and (\ref{AppendixEq:cov-simplified}) into (\ref{QUDM-def}), in order to obtain
\begin{equation}
\begin{split}
Q_{UDM} & =(\bm{\mu_{1}}-\bm{\mu_{12}})^{T}(\bm{C_{1}}-\bm{C_{12}})^{-1}(\bm{\mu_{1}}-\bm{\mu_{12}}), \\
& = [\bm{C_{1}}(\bm{C_{1}}+\bm{C_{2}})^{-1}(\bm{\mu_{1}}-\bm{\mu_{2}})]^T [\bm{C_{1}}^{-1}(\bm{C_{1}}+\bm{C_{2}})\bm{C_{1}}^{-1}] [\bm{C_{1}}(\bm{C_{1}}+\bm{C_{2}})^{-1}(\bm{\mu_{1}}-\bm{\mu_{2}})], \\
& = (\bm{\mu_{1}}-\bm{\mu_{2}})^{T}(\bm{C_{1}}+\bm{C_{2}})^{-1}(\bm{C_{1}}+\bm{C_{2}})(\bm{C_{1}}+\bm{C_{2}})^{-1}(\bm{\mu_{1}}-\bm{\mu_{2}}), \\
& =(\bm{\mu_{1}}-\bm{\mu_{2}})^{T}(\bm{C_{1}}+\bm{C_{2}})^{-1}(\bm{\mu_{1}}-\bm{\mu_{2}}).
\end{split}
\end{equation}
However, this is exactly the definition of IOI for two-experiments up to a factor of 2. This provides an exact analytical derivation of $Q_{\text{UDM}}=2\text{IOI}$ in the Gaussian and weak prior limit.

\bibliography{IOIpaperIII}

\end{document}